\documentclass[accepted]{melba}

\usepackage{amsmath,amsfonts}
\usepackage{amssymb}
\usepackage{booktabs}
\usepackage{verbatim}
\usepackage{import}
\usepackage{multirow}
\usepackage{hyperref}
\usepackage[capitalize]{cleveref}
\usepackage{caption}
\usepackage{subcaption}
\usepackage[utf8]{inputenc}
\usepackage[english]{babel}

\usepackage{csquotes}
\usepackage{adjustbox}
\usepackage{xcolor}
\usepackage{float}
\restylefloat{table}
\usepackage{placeins}
\usepackage{adjustbox, array, makecell, booktabs}
\usepackage{todonotes}

\melbaid{2025:036}  
\doi{https://doi.org/10.59275/j.melba.2025-6747}
\melbaauthors{Lee, Shi and King}  
\email{tiarna.lee@kcl.ac.uk}
\volume{3}
\firstpageno{808}  
\melbayear{2025}  
\datesubmitted{2025-04-21}  
\datepublished{2025-12-30}  

\melbaspecialissue{Special issue on Fairness of AI in Medical Imaging (FAIMI)}
\melbaspecialissueeditors{Veronika Cheplygina, Aasa Feragen, Andrew King, Ben Glocker, Enzo Ferrante, Eike Petersen, Esther Puyol-Antón, Melanie Ganz-Benjaminsen}

\ShortHeadings{Understanding-informed Bias Mitigation for Fair CMR Segmentation}{Tiarna Lee}

\title{Understanding-informed Bias Mitigation for Fair CMR Segmentation}

\author{
	\firstname Tiarna \surname Lee\aff{5}\orcid{0000-0001-5180-8884},
	\name Esther \surname Puyol-Anton\aff{5, 3} \orcid{0000-0002-9789-6629}
    \name Bram \surname Ruijsink\aff{5, 2}
    \name Pier-Giorgio \surname Masci\aff{5} \orcid{0000-0001-5196-9530}
    \name Louise  \surname Keehn\aff{4}
    \name Phil \surname Chowienczyk\aff{4} \orcid{0000-0003-4507-038X}
    \name Emily \surname Haseler\aff{4} \orcid{0000-0003-0844-686X}
    \name Miaojing \surname Shi\aff{1} \orcid{0000-0002-4933-0073}
    \name Andrew \surname King\aff{5} \orcid{0000-0002-9965-7015}
}
\affiliations{
	\num 1 \addr  College of Electronic and Information Engineering, Tongji University \\
	\num 2 \addr Guy's and St Thomas' NHS Foundation Trust \\ 
	\num 3 \addr HeartFlow Inc, United States \\
    \num 4 \addr British Heart Foundation Centre, Department of Clinical Pharmacology, King's College London \\
    \num 5 School of Biomedical Engineering \& Imaging Sciences, King's College London 
}

\abstract{
}


\begin{document}

\twocolumn[\maketitle]

\section*{Abstract}
Artificial intelligence (AI) is increasingly being used for medical imaging tasks. However, there can be biases in AI models, particularly when they are trained using imbalanced training datasets. One such example has been the strong ethnicity bias effect in cardiac magnetic resonance (CMR) image segmentation models. Although this phenomenon has been reported in a number of publications, little is known about the effectiveness of bias mitigation algorithms in this domain. We aim to investigate the impact of common bias mitigation methods to address bias between Black and White subjects in AI-based CMR segmentation models. Specifically, we use oversampling, importance reweighing and Group DRO  as well as combinations of these techniques to mitigate the ethnicity bias. Second, motivated by recent findings on the root causes of AI-based CMR segmentation bias, we evaluate the same methods using models trained and evaluated on cropped CMR images. We find that bias can be mitigated using oversampling, significantly improving performance for the underrepresented Black subjects whilst not significantly reducing the majority White subjects' performance.  
Using cropped images increases performance for both ethnicities and reduces the bias, whilst adding oversampling as a bias mitigation technique with cropped images reduces the bias further. When testing the models on an external clinical validation set, we find high segmentation performance and no statistically significant bias.

\section{Introduction}
	
Artificial intelligence (AI) is increasingly being used to aid medical diagnosis, prognosis and treatment planning. However, AI models have been shown to exhibit bias by protected attributes in many different applications \citep{Larrazabal2020, Seyyed-Kalantari2021, Klingenberg2023HigherDetection}, including AI-based segmentation of cardiac magnetic resonance (CMR) images \citep{Puyol-Anton2021FairnessSegmentation,Puyol-Anton2022FairnessSegmentation,Lee2022ASegmentation,Lee2023AnSegmentation}.
AI bias can have detrimental downstream impacts in medical imaging applications. For example, CMR segmentations are used to derive biomarkers whose values impact patient management, so greater errors in these biomarkers for certain protected groups can lead to inappropriate treatment choices and worse outcomes \citep{Puyol-Anton2022FairnessSegmentation}.

Previous work has aimed to address biases in AI models for medical imaging tasks by using generic bias mitigation methods.  For example,
\cite{Zong2022MEDFAIR:Imaging} proposed a framework with eleven algorithms which aimed to measure and mitigate biases in medical imaging classification datasets.
However, the fairness gains achieved by such methods often cause reduced performance for some protected groups, a phenomenon known as the `fairness-accuracy trade-off' \citep{Li2024TheSurvey}. Furthermore, research has suggested that, whilst such generic bias mitigation approaches may appear to reduce bias when evaluated on internal validation sets, the fairness gains often do not hold when evaluated externally \citep{Schrouff2022DiagnosingSettings, Yang2024TheGeneralization}.
We hypothesise that a possible reason for this lack of effectiveness is that these methods take a ``blind'' approach to addressing bias in AI models, i.e. they apply a generic bias mitigation algorithm that does not take account of the underlying cause of the bias. More recently, a new strand of research has emerged which attempts to understand the bias, with a view to using this knowledge to develop more informed mitigation approaches. For example, \cite{Glocker2023AlgorithmicModels} analysed bias in chest X-ray classifiers, applying test-set resampling, multitask learning, and model inspection to provide insights into the way protected attributes were encoded in the AI model. Similarly, \cite{Olesen2025SlicingMethods} used `slice discovery methods' to reveal the root causes of sex bias, also in chest X-ray classifiers. Such methods potentially enable the development of bias mitigation techniques that are targeted at the underlying cause(s) of the bias, and could lead to improved and more robust mitigation, which will hold under the effects of domain shift to external validation sets.

Recently, an investigation was performed to discover the root cause(s) of AI-based CMR segmentation bias \citep{Lee2025AnSegmentation}. One key finding of this work was that distributional differences in areas outside the heart were significant in the learning of biased representations in AI segmentation models. This finding opens up new avenues of enquiry in bias mitigation, which we explore in this paper.

\section{Related Works}

\subsection{Bias in medical imaging}
AI bias has been observed in a wide range of medical imaging modalities and tasks. For example, \cite{Seyyed-Kalantari2021} found bias in chest X-ray classification models in terms of sex and ethnicity. A CNN-based model
showed a favourable bias towards males and older people, with these groups having higher true positive rates. Similarly, \cite{Seyyed-Kalantari2021a} found that younger people, females, patients under 20 years old, Black patients and Hispanic patients had higher rates of under-diagnosis.
\cite{Larrazabal2020} also used chest X-ray data, and investigated the effect of training a thoracic disease classification model on datasets that were imbalanced by sex. They found a relationship between training set representation and performance for both sexes, and that training a model with more balanced data did not significantly decrease accuracy for the majority group but significantly improved performance for the under-represented group. 

Bias in brain magnetic resonance imaging (MRI) classification has also been reported. A key early work here was \cite{Petersen2022FeatureDetection}, who reported bias in a CNN-based model for Alzheimer's disease (AD) detection.
Similarly, \cite{Klingenberg2023HigherDetection} found higher accuracy in MRI-based  AD detection in females than males, despite balancing the training dataset by age and sex.
\cite{Wang2023BiasStudies} also assessed bias in brain MRI disease classification, investigating the impact of training choices on bias.

In dermatology imaging , \cite{Abbasi-Sureshjani2020} found bias by age and sex when training lesion classification models.
\cite{Daneshjou2022DisparitiesSet} found disparities in accuracy
between lighter and darker skin tones, with the model achieving higher accuracy on the lighter skin tones. Fine-tuning the model on a diverse dataset closed this gap in performance and improved overall model performance so that it was equal to, or exceeded, clinician accuracy.


\cite{Puyol-Anton2021FairnessSegmentation} was the first paper to report bias in AI-based segmentation models. This work found that an nnU-Net model trained on ethnicity-imbalanced CMR data produced biased results by ethnicity. The segmentation performance favoured White subjects, who were in the majority in the training set.
Subsequent work has investigated the downstream clinical impact of this bias \citep{Puyol-Anton2022FairnessSegmentation} and analysed it in more controlled experiments by both ethnicity and sex \citep{Lee2022ASegmentation, Lee2023AnSegmentation}. Segmentation bias has since also been reported in brain MRI \citep{Ioannou2022ASegmentation}, dermatology images \citep{Bencevic2024UnderstandingSegmentation} and orthopaedic radiographs \citep{Siddiqui2024FairHealthcare}.

\subsection{Bias mitigation methods}
Bias mitigation methods aim to reduce the bias observed between protected groups in an AI model. They can be applied as pre-, in- or post-processing methods \citep{Mehrabi2019ALearning}. Pre-processing methods aim to transform the data before training to remove bias. These methods include targeted data collection and importance reweighing (which can also be performed as an in-processing method). Efforts to collect data from more diverse populations include a dataset of dermatology data from four African countries \citep{Gottfrois2024PASSIONAfrica},
a leprosy skin imaging dataset from Brazil \citep{Barbieri2022ReimaginingData}, brain MRI images from women with fibromyalgia in Mexico \citep{Balducci2022AFibromyalgia} and brain tumour segmentation data from Nigeria \citep{Adewole2024ExpandingDataset}.

In-processing methods can be used to change the objective function or apply constraints to the model during training to reduce bias.
An example of such an approach is Group Distributionally Robust Optimisation, or Group DRO \citep{Sagawa2020DISTRIBUTIONALLYGENERALIZATION}, which alters the loss function to optimise performance for the worst performing group in a dataset.
Group DRO was one of the approaches evaluated for mitigating bias in chest X-ray classifiers in the comparative analysis by \cite{Zhang2022ImprovingClassifiers}.
Another example of a modified loss function is Pareto minimax optimisation \citep{Martinez2020MinimaxPerspective}, which aims to minimise an importance-weighted maximum `risk' across protected groups.
Adversarial learning has also been suggested as an in-processing method for bias mitigation. In \cite{Madras2018}, a model was adversarially trained to learn fair representations using an encoder-decoder network.
Similarly, \cite{Zhang2018} used adversarial debiasing with the added constraint of satisfying fairness definitions such as demographic parity, equalised odds or equalised opportunity. 

Post-processing methods aim to modify the predictions made by the model based on a subject's protected attributes. \cite{Lohia2019BiasFairness} proposed an algorithm which assesses an individual sample's prediction, establishes whether it is biased and changes the prediction to the privileged group's label if the sample experiences bias. Reject option classification, proposed in \cite{Kamiran2012DecisionClassification}, considers the confidence of predictions. For a binary classifier, prediction probabilities close to 0 or 1 represent confident predictions, whereas probabilities close to 0.5 represent more uncertain predictions. Samples are not assigned a label if their probabilities lie within a certain uncertainty range as they are considered more prone to bias and are relabelled depending on the group they belong to.

Bias mitigation in AI-based segmentation has received relatively little attention compared to classification tasks. Relevant works here include FairSeg \citep{Tian2023FairSeg:Scaling}, which published a
fairness dataset for medical image segmentation and proposed a fair error-bound scaling approach to reweight the loss function. \cite{Siddiqui2024FairHealthcare} evaluated stratified batch sampling, a balanced dataset model and a protected group-specific model for orthopaedic image segmentation. Finally, in CMR segmentation, \cite{Puyol-Anton2021FairnessSegmentation} internally evaluated protected group-specific models, oversampling of minority protected groups and a multi-task learning approach which learnt both a segmentation model and a protected attribute classifier.

\section{Contributions}
In the context of AI-based segmentation of cine CMR, the main contributions of this novel work are:
\begin{enumerate}
    \item We perform the most extensive and comprehensive investigation to date of multiple bias mitigation methods in CMR segmentation, as well as combinations of these methods.
    \item We perform one of the first investigations into using knowledge of the root cause of bias for mitigation. Specifically, we train AI segmentation models using CMR images that are cropped to remove features outside of the heart. We evaluate a baseline model and state-of-the-art bias mitigation techniques in this setting.
    \item We evaluate the efficacy of all approaches under both internal and external validation settings.
\end{enumerate}
A preliminary version of this work has been published in \cite{Lee2025DoesSegmentation}. This paper extends that work by (i) incorporating a wider range of metrics and more in-depth analysis and discussion, (ii) inclusion of a new clinically-applicable `cascaded' cropping based mitigation approach and (iii) including external validation of all approaches on a clinical dataset.

\section{Materials and Methods}

\subsection{Data}

To train and internally validate all models we used CMR images from the UK Biobank \citep{Peterson2016}. The dataset consists of end diastolic (ED) and end systolic (ES) cine short-axis images from 5,778 subjects. Manual segmentation of the left ventricular blood pool (LVBP), left ventricular myocardium (LVM), and right ventricular blood pool (RVBP) was performed for the ED and ES images of each subject. The LV endocardial and epicardial borders and the RV endocardial border were outlined using cvi42 (version 5.1.1, Circle Cardiovascular Imaging Inc., Calgary, Alberta, Canada). The same guidelines were provided to a panel of ten experts with one expert annotating each image. Each expert was provided with a random selection of images for annotation which included subjects of different sexes and ethnicities. They were not provided with demographic information about the subjects.

Previous work \citep{Lee2022ASegmentation, Lee2023AnSegmentation} has shown that bias is greater when the imbalance between ethnicities in the training set is greater. Therefore, we curated a dataset where biases would be significant to allow for better evaluation of mitigation methods. Our main training set comprises 15 Black subjects and 4,221 White subjects, all randomly sampled from the full dataset. The remaining subjects were used as the internal validation test set. We also investigate the effect of using different proportions of Black and White subjects in the training set in the Supplementary Material. The demographic information of the subjects in the training and test sets can be seen in \cref{tab:addressing_bias_training_data} and \cref{tab:addressing_bias_test_data}. Note that training was performed using only Black and White subjects but testing was performed using all other available subjects including Mixed, Asian, Chinese and Other. All ethnicity information refers to self-reported ethnicity based on the categories provided in the UK Biobank dataset.

In addition, for external validation a dataset of cine short-axis CMR images from St. Thomas' Hospital, London was used. All subjects were scanned using a 1.5T MRI scanner (Aera-Magneton, Siemens Healthcare, Erlangen, Germany) between November 2019 and November 2021. The patients scanned had suspected hypertension. Ethnicity was recorded as Black if both parents self-identified as African descendants or White if both parents self-identified as European descendants. The demographic information for the subjects used can be seen in \cref{tab:addressing_bias_external_test_data}. 
Ground truth segmentations were performed manually by clinical experts using cvi42 (Circle Cardiovascular Imaging Inc., Calgary, Alberta, Canada). Further details of the imaging protocol can be found in \cite{Georgiopoulos2024EthnicityResonance}.



\begin{table}
\centering
\caption{Characteristics of subjects used in the training dataset. Mean (standard deviation) values are presented for each characteristic. Statistically significant differences between subject groups and the overall average are indicated with an asterisk * (p $<$ 0.05) and were determined using a two-tailed Student’s t-test.}
\begin{adjustbox}{width=0.8\columnwidth,center}
\begin{tabular}{|c|c|c|c|}
\hline
\textbf{Health measure}       & \textbf{Overall}     & \textbf{White}       & \textbf{Black}                            \\
\hline n subjects          & 4236         & 4221         & 15                              \\
\hline Age (years)           & 64.6 (7.7)  & 64.6 (7.7)  & 57.0 (4.9)*                       \\
\hline Weight (kg) & 77.0 (15.0) & 77.0 (15.0) & 74.8 (11.9)* \\
\hline Height (cm) & 169.5 (9.2) & 169.5 (9.2) & 168.8 (6.9)*                      \\
\hline Body Mass Index      & 26.7 (4.3)  & 26.7 (4.3) & 26.2 (3.7)  \\ \hline
\end{tabular}
\label{tab:addressing_bias_training_data}
\end{adjustbox}
\end{table}


\begin{table*}[ht]
\centering
\caption{Characteristics of subjects used in the internal validation test dataset. Mean (standard deviation) values are presented for each characteristic. One-way ANOVA was used to test for differences across ethnic groups. Statistically significant differences between groups are indicated with an asterisk. * (p $<$  0.05).}
\begin{adjustbox}{width=\textwidth,center}
\label{tab:addressing_bias_test_data}
\begin{tabular}{|c|c|c|c|c|c|c|c|}
\hline
Health measure & Overall & White & Mixed & Asian & Black & Chinese & Other \\ \hline

n subjects & 1542 & 469 & 170 & 387 & 223 & 111 & 182 \\
 Age (years) *& 61.7 (7.9) & 64.8 (7.7) & 59.8 (7.2) & 61.0 (8.2) & 59.1 (7.1) & 59.7 (6.4) & 62.1 (7.5) \\
Height (cm) *& 167.4 (9.2) & 169.8 (9.4) & 166.4 (8.5) & 166.7 (8.6) & 168.7 (9.4) & 162.4 (7.1) & 165.4 (9.7) \\
Weight (kg) *& 75.1 (15.1) & 77.7 (15.1) & 75.1 (15.3) & 72.3 (12.8) & 82.0 (16.0) & 63.5 (9.9) & 73.3 (15.5) \\
Body Mass Index *& 26.7 (4.4) & 26.9 (4.2) & 27.1 (5.2) & 25.9 (3.7) & 28.8 (5.1) & 24.0 (3.1) & 26.6 (4.4) \\ \hline

\end{tabular}
\end{adjustbox}

\end{table*}

\begin{table}[H]
\centering
\caption{Characteristics of subjects used in the external validation test set. Mean (standard deviation) values are presented for each characteristic. Statistically significant differences between subject groups and the overall average are indicated with an asterisk * (p $<$ 0.05) and were determined using a two-tailed Student’s t-test.}
\begin{adjustbox}{width=0.8\columnwidth,center}
\begin{tabular}{|c|c|c|c|}
\hline
\textbf{Health measure}       & \textbf{Overall}     & \textbf{White}       & \textbf{Black}                            \\
\hline n subjects          & 84         & 30         & 54  \\
\hline Age (years)           &  46.0 (12.8)& 40.5 (14.1)& 49.1 (11.0)                   \\
\hline Weight (kg) &  94.6 (19.7)& 92.6 (19.3) &95.1 (20.2)\\
\hline Height (cm) & 174.3 (8.96)&  176.9 (7.33)   & 172.9 (9.48)          \\
\hline Body Mass Index  &  31.2 (6.23)& 29.6 (6.13) & 31.8 (6.23)\\ \hline
\end{tabular}
\label{tab:addressing_bias_external_test_data}
\end{adjustbox}
\end{table}

\subsection{Baseline Model}

As a baseline model, we trained a 2D nnU-Net v1 model \citep{Isensee2020} using the UK Biobank training set to segment the LVM, LVBP and RVBP. 

\subsection{Oversampling}

We also trained a 2D nnU-Net model using the same data as the baseline but applied oversampling during batch selection \citep{Kamiran2012DataDiscrimination}. Oversampling refers to the process of increasing the sampling of a minority group in a dataset. Here, we oversample the Black subjects in the training set so that they were equal to the number of White subjects in each batch used during training. This was performed using random sampling with replacement so each subject could in principle be selected more than once in a training batch.

\subsection{Reweighing}

An nnU-Net model was also trained using a reweighing mitigation strategy \citep{Kamiran2012DataDiscrimination}. Reweighing refers to the process of increasing the importance of under-represented groups to the model.
We implemented this strategy 
by adding a weighting term to the combined Cross Entropy (CE)-Dice loss function of the nnU-Net. Each group was weighted inversely proportionally to the group size, as shown in \cref{reweighting1}. The weights were then normalised so that they summed to 1, as shown in  \cref{reweighting2}.


\begin{align}
\text{Weights per group: } &\quad w_g = \frac{n_G}{n_g + \epsilon}, \quad \text{where } \epsilon = 10^{-6}. \label{reweighting1} \\ 
\text{Normalized weights: } &\quad \hat{w}_g = \frac{w_g}{\sum_{j=1}^{N_G} w_j}, \quad g = 1, 2, \dots, N_G.
\label{reweighting2}
\end{align}

\noindent where $n_g$ is the number of samples in protected group $g$, $n_G$ is the number of samples in all groups and $N_G$ is the number of groups. 

\subsection{Group Distributionally Robust Optimisation}

The final mitigation approach was Group DRO, which was first proposed in \cite{Sagawa2020DISTRIBUTIONALLYGENERALIZATION}. The method aims to optimise the performance of the worst-performing group in a dataset.
The Group DRO loss function can be formalised as:

\begin{equation}
{L_{DRO}} = max_{g \in G}\frac{1}{{{n_g}}}\sum\nolimits_{i \in g} {L\left(y_i, \hat{y_i}\right)} \tag{2}
\end{equation}

\noindent where $L$ is the loss function computed between predicted labels $\hat{y_i}$ and ground truth labels $y_i$.


In this method, CE loss was used instead of CE-Dice loss which was used for the oversampling and reweighing experiments. The reason for this is that Group DRO uses losses from individual samples to calculate the average loss for the groups. However, Dice loss is calculated using global statistics of the true positives, false positives and false negatives for a group or batch. It is non-additive as the numerator and denominator will change if the calculation is performed on a per-sample basis rather than for a group or batch. For Dice loss, the average group loss and global loss are different, which causes instability in training.

\subsection{Training Using Combinations of Mitigation Methods}

We also combined the mitigation methods into pairs to test whether combinations of methods would improve performance. This results in three additional methods: oversampling + Group DRO, reweighing + Group DRO, and oversampling + reweighing. These methods were applied in the same way as above. CE loss was used for experiments with Group DRO and CE-Dice loss was used for oversampling + reweighing.

\subsection{Training Using Cropped Images}
\label{sec:image_cropping}

Following the findings of \cite{Lee2025AnSegmentation}, which found that areas outside the heart were a contributing factor to CMR segmentation performance bias, we also performed experiments using all of the above techniques for a nnU-Net model trained using cropped CMR images. The images were cropped around the heart using a bounding box defined based on a segmentation mask. All images were cropped to the same size, i.e. the size of the largest heart in the training set plus a buffer of 5 pixels in both the $x$ and $y$ directions.

When using cropped images, we evaluated two different approaches. First, we used the ground truth segmentations to calculate the cropping region at both training and inference time. Note that such a technique could not be used when the model is deployed since ground truth segmentations would not be available at inference time. Therefore, we evaluate this method to establish an upper bound on the performance of the cropping-based mitigation approach.

\begin{figure*}[!ht]
    \centering
    \includegraphics[width=\textwidth]{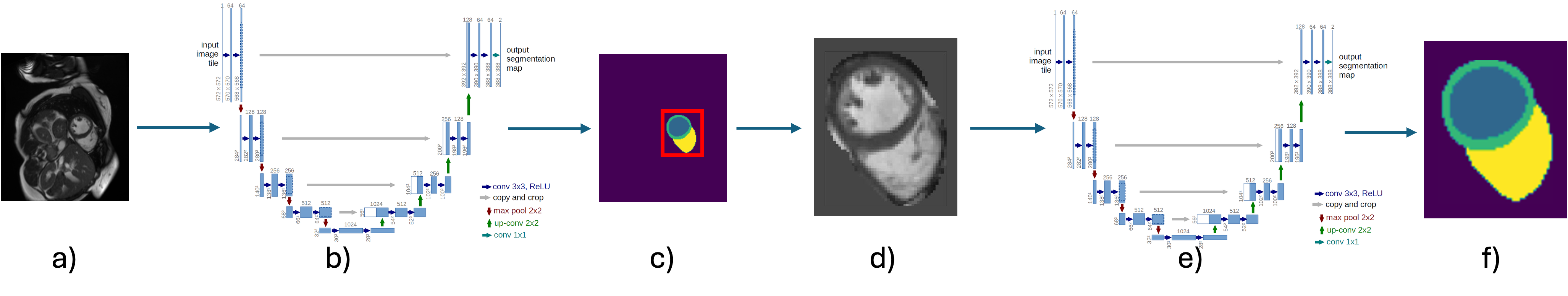}
    \caption{`Cascaded' approach to bias mitigation based on using cropped images. Images (a) are first segmented using a full-image nnU-Net (b) to produce a segmentation (c). This segmentation is then used to crop the images (d) and used in another nnU-Net model trained using ground truth segmentation-based cropped images (e) to produce the final segmentations (f).}
    \label{fig:cascaded_figure}
\end{figure*}

Second, we developed a `cascaded' cropping-based approach,
in which the cropping region was estimated using an initial nnU-Net-based segmentation. This first nnU-Net model was trained using the full images and its output was used to estimate the cropping region. The resulting cropped image was then used as input to a second nnU-Net model which was trained using cropped images. The croppings used on the training data in this second nnU-Net were based the ground truth segmentations, which are available at training time. This approach is illustrated in \cref{fig:cascaded_figure}. 




\subsection{Evaluation}
For each of the methods described above, performance was measured by finding the overall Dice similarity coefficient (DSC) and Hausdorff distance (HD) for subjects in the internal and external validation test sets.  Performance was quantified using the median and inter-quartile range (IQR) of the DSC/HD. Unless otherwise stated, we report the values of these metrics averaged over the LVBP, LVM and RVBP.

Fairness metrics are also reported. The fairness gap (FG), represents the difference in median DSC between the protected groups (i.e. ethnicities) and is given by $FG = D_{White} - D_{Black}$ where $D$ is the median DSC. The skewed error ratio (SER),  as defined in \cite{Puyol-Anton2021FairnessSegmentation},  is given by  $ SER = \frac{max_g(1-D_{g})}{min_g(1-D_{g})}$ where $D_g$ is the median DSC for protected group $g$.  This measures the ratio of the errors between the median DSCs for the worst-performing and best-performing protected groups. For both fairness metrics a low value represents less bias and a more fair model will have a FG closer to 0 and a SER closer to 1.

\subsection{Code availability}
{\sloppy

The code to reproduce this work is available at 
\url{https://github.com/tiarnaleeKCL/nnUNet-bias-mitigation}.

\par}

\section{Results}

\subsection{Internal validation}

\begin{figure*}[!ht]
\centering
\begin{subfigure}{.45\linewidth}
    \centering
    \includegraphics[scale=0.6]{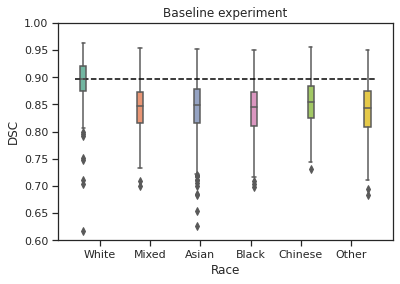}
    \caption{}\label{fig:Baseline_bias_mitigation}
\end{subfigure}
    \hfill
\begin{subfigure}{.45\linewidth}
    \centering
    \includegraphics[scale=0.6]{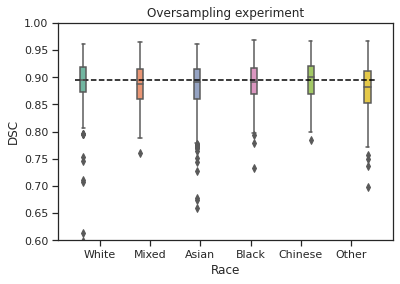}
    \caption{}\label{fig:Resampling}
\end{subfigure}

\begin{subfigure}{.45\linewidth}
    \centering
    \includegraphics[scale=0.6]{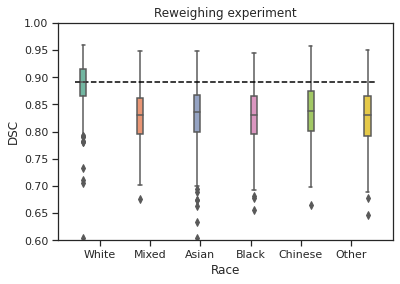}
    \caption{}\label{fig:Reweighing}
\end{subfigure}
    \hfill
\begin{subfigure}{.45\linewidth}
    \centering
    \includegraphics[scale=0.6]{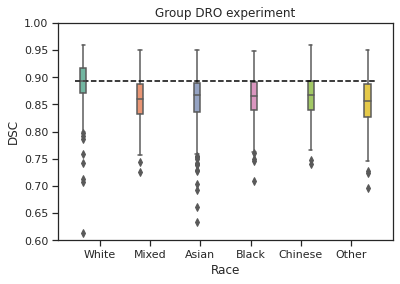}
    \caption{}\label{fig:GroupDRO}
\end{subfigure} 

\begin{subfigure}{.45\linewidth}
    \centering
    \includegraphics[scale=0.6]{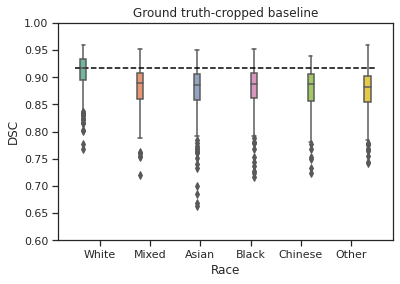}
    \caption{}\label{fig:internal_GT_cropped_baseline}
\end{subfigure}
    \hfill
\begin{subfigure}{.45\linewidth}
    \centering
    \includegraphics[scale=0.6]{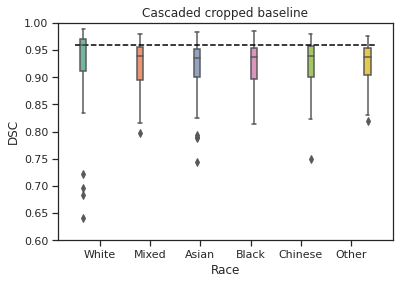}
    \caption{}\label{fig:internal_cascaded_cropped_baseline}
\end{subfigure} 

\caption{Overall DSC for bias mitigation methods on uncropped images. The dashed line indicates median DSC for White test subjects.}

\label{fig:Bias_mitigation}
\end{figure*}

The internal validation results of the baseline and bias mitigation methods can be seen in \cref{tab:Bias_mitigation_results} and \cref{fig:Bias_mitigation}. Statistical tests were performed using Mann-Whitney U tests on the overall DSC scores. Oversampling was the only method to increase performance for the Black subjects such that there was no significant difference between the median DSC scores of the Black and White subjects. The method increased the median DSC for Black subjects by 0.045. Fairness performance metrics (SER and FG) also decreased, showing more equitable performance. Using oversampling also caused performance to increase for the other ethnicities compared to the baseline (see \cref{fig:Bias_mitigation}). This is perhaps surprising as performance on these other ethnicities was not optimised during training. Further results on the effect of different levels of oversampling, both without and without cascaded cropping, can be seen in \cref{fig:changing_oversampling}. Plots of performance (median DSC) against FG for these two approaches can be seen in \cref{fig:performance_vs_fairness}. The effect of using different proportions of Black and White subjects in the training set are reported in \cref{fig:changing_black_subjects}. Additional experiments using Asian and White subjects can be seen in \cref{tab:bias-mitigation-results-aw}
. 

The other mitigation methods did not significantly improve performance for the Black subjects. Reweighing resulted in worse performance for the Black subjects, with the median DSC decreasing and HD increasing. Group DRO resulted in increased performance for the Black subjects but performance remained significantly lower than for the White subjects. Both oversampling and Group DRO slightly decreased median DSC for White subjects (although not statistically significantly), but reweighing significantly decreased median DSC for White subjects.

\begin{table}[!ht]    
\centering
\caption{DSC and HD values for each of the bias mitigation methods tested on the internal validation set. The p-values were computed between White and Black subjects based on a two-sided Mann Whitney U test  on the DSC scores. * p$<$ 0.05. The best median DSC score for Black and White subjects is shown in bold.}
\begin{adjustbox}{width=\columnwidth,center}

\begin{tabular}{|c|c|c|c|c|c|c|c|c|}
\hline
& \multicolumn{2}{c}{Baseline *} & \multicolumn{2}{|c|}{\begin{tabular}[c]{@{}c@{}}Oversampling \\ p = 0.22\end{tabular}} & \multicolumn{2}{c}{Reweighing *} & \multicolumn{2}{|c|}{Group DRO *} \\
\hline
& White  & Black  & White  & Black  & White  & Black  & White  & Black \\
\hline
Median DSC & \textbf{0.896}  & 0.846  & 0.894  & \textbf{0.891 } & 0.891  & 0.831  & 0.893  & 0.865 \\

IQR DSC & 0.046 & 0.064 & 0.045 & 0.049 & 0.049 & 0.069 & 0.047 & 0.051 \\

Median HD (mm) & 6.725 & 9.364  & 6.835 & 7.164 & 7.174 & 10.037  & 6.802 & 8.286\\
IQR HD (mm) & 3.707 & 4.725 & 3.713 & 3.861 & 4.207 & 4.902 & 3.581 & 4.157 \\

SER     &\multicolumn{2}{c}{1.486}  &\multicolumn{2}{|c|}{1.032}  &\multicolumn{2}{c}{1.544}        &\multicolumn{2}{|c|}{1.260}\\
Fairness gap    &\multicolumn{2}{c}{0.050}  &\multicolumn{2}{|c|}{0.003}  &\multicolumn{2}{c}{0.059}  &\multicolumn{2}{|c|}{0.028}\\\hline
\end{tabular}
\label{tab:Bias_mitigation_results}
\end{adjustbox}
\end{table}

Combining the mitigation methods did not produce significantly less biased results, as shown in \cref{tab:combined_bias_mitigation}. Mann-Whitney U tests were performed to test the significance of differences between the overall DSC scores. All three combinations decreased performance for White subjects but increased performance for Black subjects. The best combination was oversampling and Group DRO which reduced the performance for the White subjects the least and improved performance for the Black subjects the most, leading to the lowest FG.

\begin{table}[!ht]
\centering
\caption{DSC and HD values for the combined bias mitigation methods tested on the internal validation set using original sized images. The p-values were computed between White and Black subjects based on a two-sided Mann Whitney U test on the DSC scores. * p$<$ 0.05. The best median DSC score for Black and White subjects is shown in bold.}

\begin{adjustbox}{width=\columnwidth,center}

\begin{tabular}{|c |c |c |c |c |c |c |c |c|}
\hline
 & \multicolumn{2}{c|}{\begin{tabular}[c]{@{}c@{}}Baseline \\ *\end{tabular}} 
 & \multicolumn{2}{c}{\begin{tabular}[c]{@{}c@{}}Oversampling \\+ Group DRO \\ *\end{tabular}}
 & \multicolumn{2}{|c|}{\begin{tabular}[c]{@{}c@{}}Group DRO \\+ Reweighing \\ *\end{tabular}}
 & \multicolumn{2}{c|}{\begin{tabular}[c]{@{}c@{}}Oversampling \\+ Reweighing \\ *\end{tabular}} \\
 \hline

 & White & Black & White & Black & White & Black & White & Black \\ \hline

Median DSC & \textbf{0.896} & 0.846 & 0.889 & \textbf{0.881} & 0.893 & 0.863 & 0.889 & 0.871 \\

IQR DSC& 0.046 & 0.064 & 0.048 & 0.054 & 0.047 & 0.055  & 0.051 & 0.057 \\

Median HD (mm) & 6.725 & 9.364 & 6.985 & 7.740  & 6.831 & 8.417 & 7.364 & 8.654  \\
IQR HD (mm) & 3.707 & 4.725 & 3.666 & 4.079 &3.520 & 4.001 & 4.320 & 4.558 \\

SER & \multicolumn{2}{c}{1.486} & \multicolumn{2}{|c|}{1.08} & \multicolumn{2}{c}{1.285} & \multicolumn{2}{|c|}{1.161} \\
Fairness gap & \multicolumn{2}{c}{0.050} & \multicolumn{2}{|c|}{0.009} & \multicolumn{2}{c}{0.031} & \multicolumn{2}{|c|}{0.017} \\\hline

\end{tabular}
\label{tab:combined_bias_mitigation}
\end{adjustbox}
\end{table}

As shown in \cref{tab:Bias_mitigation_cropped_results}, the baseline model trained using cropped images improved performance for both White and Black subjects and reduced bias. Oversampling significantly improved performance for the Black subjects compared to the baseline, resulting in performance that was higher than for White subjects. Group DRO improved the DSC for Black subjects compared to the baseline but this increase was not significant.  However, the HD for both groups of subjects decreased. Reweighing made performance slightly worse for both ethnicities, with median DSC decreasing and SER and FG measures increasing. 

\begin{table}[!ht]
\centering
\caption{DSC and HD values for each of the bias mitigation methods tested on the internal validation set using cropped images. The p-values were computed between White and Black subjects based on a two-sided Mann Whitney U test on the DSC scores. The best median DSC score for Black and White subjects is shown in bold.}

\begin{adjustbox}{width=\columnwidth,center}
\begin{tabular}{@{}|c|cc|cc|cc|cc|@{}|}
\hline
&\multicolumn{2}{c}{Baseline *}&\multicolumn{2}{|c|}{Oversampling *}&\multicolumn{2}{c}{Reweighing *}&\multicolumn{2}{|c|}{Group DRO *}\\ \hline
&White  &Black  &White  &Black  &White  &Black  &White  &Black\\\hline

Median DSC & 0.917 & 0.888 & 0.906 & \textbf{0.914} & 0.915 & 0.877  & \textbf{0.917} & 0.903\\

IQR DSC  & 0.039 & 0.047 & 0.031 & 0.044  & 0.036 & 0.052  & 0.035 & 0.046\\

Median HD (mm)  & 5.713 & 7.895 & 6.394 & 6.479  & 5.987 & 8.945  & 5.700 & 7.170\\
IQR HD (mm) & 2.669 & 3.960 & 3.359 & 3.211 & 3.073 & 4.097 & 2.805 & 3.584\\

SER     &\multicolumn{2}{c|}{1.340} &\multicolumn{2}{c|}{1.088} &\multicolumn{2}{c|}{1.448} &\multicolumn{2}{c|}{1.161}\\
Fairness gap    &\multicolumn{2}{c|}{ 0.028} &\multicolumn{2}{c|}{-0.008} &\multicolumn{2}{c|}{0.038}  &\multicolumn{2}{c|}{0.013}\\\hline
\end{tabular}
\label{tab:Bias_mitigation_cropped_results}
\end{adjustbox}
\end{table}

\cref{tab:cascaded_cropping_internal} and \crefrange{fig:internal_GT_cropped_baseline}{fig:internal_cascaded_cropped_baseline} show the results for the cascaded cropping on the internal validation dataset. Surprisingly, using the cascaded cropping approach significantly improved the DSC score compared to using the ground-truth cropped results shown in \cref{tab:Bias_mitigation_cropped_results} for all bias mitigation methods. We discuss a possible reason for this in \cref{sect:discussinternal}. Using oversampling on the cascaded cropping approach did not improve DSC score but decreased the HD, SER and FG compared to the baseline. Reweighing also improved the SER and FG. 

\begin{table}[!ht]    
\centering
\caption{DSC and HD values for each of the bias mitigation methods using the cascaded cropping approach on the internal validation data. The p-values were computed between White and Black subjects based on a two-sided Mann Whitney U test on the DSC scores. * p$<$ 0.05. The best median DSC score for Black and White subjects is shown in bold.}
\begin{adjustbox}{width=\columnwidth,center}

\begin{tabular}{|c|c|c|c|c|c|c|c|c|}
\hline
&\multicolumn{2}{c}{Baseline *}&\multicolumn{2}{|c|}{Oversampling *}&\multicolumn{2}{c}{Reweighing *}&\multicolumn{2}{|c|}{Group DRO *}\\ \hline
& White  & Black  & White  & Black  & White  & Black  & White  & Black \\
\hline
Median DSC & \textbf{0.959} & 0.937 & 0.958 & 0.941 & 0.951 & 0.935 & 0.955 & \textbf{0.943}\\
IQR DSC  & 0.060 & 0.057 & 0.064 & 0.058 & 0.067 & 0.070 & 0.068 & 0.058\\

Median HD (mm) & 3.757 & 6.564  & 3.731 & 4.929 & 5.176 & 6.426 & 3.841 & 4.623\\

IQR HD (mm) & 2.268 & 5.012  & 2.129 & 2.717 & 3.262 & 4.172 & 2.039 & 3.327\\
SER     &\multicolumn{2}{c}{1.517}  &\multicolumn{2}{|c|}{1.380}  &\multicolumn{2}{c}{1.332}        &\multicolumn{2}{|c|}{1.276}\\
Fairness gap    &\multicolumn{2}{c}{0.021}  &\multicolumn{2}{|c|}{0.016}  &\multicolumn{2}{c}{0.016}  &\multicolumn{2}{|c|}{0.012}\\\hline
\end{tabular}
\label{tab:cascaded_cropping_internal}
\end{adjustbox}
\end{table}

\subsection{External validation}

\cref{tab:Overall_bias_mitigation_external} shows the external validation results.  \crefrange{tab:LVBP_results}{tab:RVBP_results} show the results broken down by cardiac region. As oversampling was found to be the best bias mitigation on internal validation, only the results for this model and the baseline model are shown. Note that, as the LVM is not routinely segmented for the ES frame in clinical workflows at our institution, results for the external validation dataset are shown for the ED frames only. 

\begin{table}[!ht]    
\centering
\caption{DSC and HD values for the baseline and oversampling methods tested on the external validation set. The p-values were computed between White and Black subjects based on a two-sided Mann Whitney U test on the DSC scores. * p$<$ 0.05. The best median DSC score for Black and White subjects is shown in bold.}
\begin{adjustbox}{width=\columnwidth,center}

\begin{tabular}{|c|c|c|c|c|c|c|c|c|}
\hline
&  \multicolumn{2}{c|}{\begin{tabular}[c]{@{}c@{}}Baseline \\ p = 0.093 \end{tabular}} & \multicolumn{2}{|c|}{\begin{tabular}[c]{@{}c@{}}Oversampling \\ p = 0.16 \end{tabular}}  \\
\hline
& White  & Black  & White  & Black  \\
\hline
Median DSC & \textbf{ 0.886} & \textbf{ 0.879} & 0.885 & 0.872 \\

IQR DSC&  0.030 & 0.043  & 0.037 & 0.038 \\

Median HD (mm) & 5.728 & 5.913 & 6.152 & 5.919  \\

IQR HD (mm) & 1.522 & 1.966 & 1.238 & 1.840  \\

SER     &\multicolumn{2}{c}{1.064}  &\multicolumn{2}{|c|}{1.106}  \\
Fairness gap    &\multicolumn{2}{c}{0.007}  &\multicolumn{2}{|c|}{0.012}  \\\hline
\end{tabular}
\label{tab:Overall_bias_mitigation_external}
\end{adjustbox}
\end{table}

Overall, the model had high performance on the external validation set. The performance was comparable to that of the bias mitigation methods tested on the internal validation set, with the baseline performance for the Black subjects being higher than for the Black subjects in the internal validation set. The fairness gap and SER were also smaller than for the internal validation set. However, for the external validation set, using oversampling decreased performance for both groups, with the FG, SER and median HD increasing. 

The approach based on cropping using ground truth segmentations produced a significantly better baseline performance in the external validation set for both groups compared to not using cropping, as shown in \cref{tab:overall_cascaded_results}.
Using oversampling slightly decreased the performance for both groups and resulted in a significantly different performance for Black and White subjects.

The results from using the cascaded cropping based approach can be seen in \cref{tab:overall_cascaded_results}. Mann-Whitney U tests were used to compare the overall DSC scores. Furthermore, the errors in the sizes of the bounding boxes using the cascaded approach compared to those calculated using the ground-truth segmentations can be seen in \cref{tab:differences_in_bounding_box}. \crefrange{tab:LVBP_cascaded_results}{tab:RVBP_cascaded_results} show the results broken down by cardiac region for the cascaded approach. Using this approach, performance was lower than the upper bound of using ground truth segmentations for cropping for both groups but performance remained higher than for the baseline model using uncropped images seen in \cref{tab:Overall_bias_mitigation_external}. Performance increased for both groups when using oversampling and the FG decreased. \cref{tab:overall_cascaded_results_sex_and_age} shows the results for the oversampling cascaded model broken down by sex and age.




\begin{table*}[!ht]
\centering
\caption{DSC and HD values for the models using cropping on the external validation dataset. The p-values were computed between White and Black subjects based on a two-sided Mann Whitney U test on the DSC scores. * p$<$ 0.05. The best median DSC score for Black and White subjects is shown in bold.}
\begin{adjustbox}{width=\textwidth,center}
\begin{tabular}{@{}|c|cc|cc|cc|cc|cc|cc|}
\hline
& \multicolumn{2}{c|}{\begin{tabular}[c]{@{}c@{}}Cropped baseline \\ ground-truth model \\ p = 0.069 \end{tabular}} & \multicolumn{2}{c|}{\begin{tabular}[c]{@{}c@{}}Cropped oversampling \\ ground-truth model \\ * \end{tabular}} & \multicolumn{2}{c}{\begin{tabular}[c]{@{}c@{}}Cropped baseline \\ cascaded model \\ p = 0.12 \end{tabular}}& \multicolumn{2}{|c|}{\begin{tabular}[c]{@{}c@{}}Cropped oversampling \\ cascaded model \\ p = 0.68\end{tabular}}\\ \hline
 & White  & Black  & White  & Black &White  &Black  &White  &Black \\\hline

Median DSC & \textbf{0.931} & 0.916 & 0.918 & 0.906 & 0.917 & 0.903 & 0.928 & \textbf{0.920}\\
IQR DSC  & 0.043 & 0.037 & 0.040 & 0.033 & 0.045 & 0.038 & 0.039 & 0.026  \\
Median HD (mm) & 4.087 & 4.090 & 4.793 & 4.594 & 4.772 & 4.954  & 4.658 & 3.875\\ 
IQR HD (mm) & 1.630 & 1.178 & 1.321 & 1.529 & 3.124 & 1.919  & 2.69 & 1. 432\\
SER   &\multicolumn{2}{c}{1.207}        &\multicolumn{2}{|c|}{1.141}  &\multicolumn{2}{c|}{1.168} &\multicolumn{2}{c|}{1.118} \\
Fairness gap  &\multicolumn{2}{c}{0.014}  &\multicolumn{2}{|c|}{0.012}  &\multicolumn{2}{c|}{0.014} &\multicolumn{2}{c|}{0.008} \\\hline
\end{tabular}
\label{tab:overall_cascaded_results}
\end{adjustbox}
\end{table*}

\section{Discussion}

\subsection{Internal validation}
\label{sect:discussinternal}

This work has performed a comprehensive examination of bias mitigation methods for AI segmentation models used for cine CMR images. We have shown that bias in CMR segmentation models can be mitigated by using such methods. In particular, oversampling minority subjects reduces bias so that there is no significant difference between the performance of the Black and White subjects. Although oversampling did not add any extra information to the dataset, the method allowed the network to train on Black subjects more frequently than if oversampling was not used, allowing for better balance between protected groups. It could be anticipated that training using a small number of (oversampled) Black subjects would increase the risk of overfitting to those subjects, having a detrimental effect on generalisation, but this effect was not seen in our experiments as test performance remained high. Previous work in \cite{Lee2025AnSegmentation} has showed that ethnicity can be classified from cine CMR images, indicating that there are distinct features in the images of different ethnicities that are recognisable to AI models. Using oversampling will allow the network to see more of these distinct features to learn better representations for the under-represented group. 

Reweighing did not improve segmentation performance, instead decreasing performance for both protected groups. This may be due to increased importance being given to a small group of subjects, decreasing focus on the larger group. As described in \cite{Lee2022ASegmentation} and \cite{Lee2023AnSegmentation}, when White subjects comprised 75\% of the training set, their segmentation performance was still lower than for the Black subjects who comprised 25\% of the training set. This suggests that segmentation of the White subjects' images may be a more difficult task as there may be more outliers and variation in the hearts than in the Black subjects. Reweighing these White subjects so that their importance is lower in the loss function could decrease accuracy. 

The combination of Group DRO and oversampling produced a model which improved performance for Black subjects but was still significantly lower than for White subjects. Therefore, using a combined method was better than using Group DRO alone but worse than oversampling alone.

Using cropped images increased performance and reduced bias compared to using uncropped images. Using oversampling with cropped images reduced bias further, but interestingly not to the same extent as using uncropped images. The cascaded cropping approach resulted in a performance that was higher than using ground truth segmentations for cropping. This may be due to the intermediate predictions on the uncropped images (c in \cref{fig:cascaded_figure}) being less accurate and the segmented area being larger than the ground truth segmentations, resulting in a bounding box which was larger and contained more information. This suggests that performance could be optimised further by increasing the size of the pixel buffer used in the cropping process.

\subsection{External validation}

Using the bias mitigation models on the external validation set produced comparable DSC scores to internal validation results. No significant bias was observed in the baseline model for this dataset, as shown in \cref{tab:Overall_bias_mitigation_external}. This is surprising as the dataset is out-of-distribution and some distributional shift is expected, which could lead to worse performance. However, unlike the models tested on the internal validation set, oversampling did not decrease the FG or improve the performance of the models under external validation. This is consistent with previous work that has reported limited effectiveness of generic bias mitigation algorithms under complex domain shifts \citep{Schrouff2022DiagnosingSettings, Yang2024TheGeneralization}.

Cropping the images increased performance overall, both using the ground truth-based cropping and the cascaded approach. Importantly, using the cascaded approach, performance was higher than the baseline model using uncropped images for both groups under external validation. Interestingly, using oversampling in combination with the cascaded approach improved performance for both groups further. These results indicate that, not only does cropping improve the performance of models for both protected groups (i.e. ethnicities), but also that ground truth segmentations are not needed to crop these images. This finding can be vital for the clinical translation of bias mitigation algorithms in CMR analysis, as better segmentation performance will allow for better assessment of clinical biomarkers and better treatment planning, prognosis and diagnosis. Provided that there is both a network trained on uncropped images and a network trained on cropped images, the method will be deployable and scalable. In the future, the first model used to localise the cardiac region of interest (b in \cref{fig:cascaded_figure}) could be replaced with a similar network trained for bounding box detection such as that proposed in \cite{He2017MaskR-CNN}.

The best method overall (cascaded cropping used with oversampling) was based on an understanding of the root cause of bias in AI-based CMR analysis as reported in \cite{Lee2025AnSegmentation}. 
Specifically, in \cite{Lee2025AnSegmentation} it was reported that the main source of the distributional shift between ethnicities (and hence the bias in segmentation performance) was outside the heart. The cascaded cropping approach is a simple but elegant approach to removing the source of the bias in a robust way but maintaining (and even improving) segmentation performance.
This important result highlights that, when using a bias understanding-informed mitigation approach in the context of CMR segmentation, there is no fairness-accuracy trade-off. 
The fairness-accuracy trade-off is sometimes used as a reason not to implement bias mitigation algorithms in clinical practice, due to the medical principle of non-maleficence. Our work suggests that this trade-off is not inevitable, but that it can be avoided by a careful analysis of bias in individual scenarios rather than application of a generic mitigation algorithm.

\subsection{Limitations and future work}

This work has some limitations. For example, only two ethnicities (Black/White and Asian/White) were used during training in our experiments. Future work could investigate the effect of mitigating biases using multiple ethnicities, in other protected attributes such as age and socioeconomic status, or in intersectional groups. There was also a relatively small number of subjects in the external dataset when compared to the number of subjects in the training and internal validation sets from the UK Biobank, so externally validating on a larger, more diverse dataset would be beneficial. 
\textcolor{blue}{In future work we also plan to investigate more advanced baseline bias mitigation methods than the ones used in this work. However, it should be noted that the literature on bias mitigation in segmentation (e.g. \cite{Puyol-Anton2021FairnessSegmentation, Bencevic2024UnderstandingSegmentation, Siddiqui2024FairHealthcare}) is limited compared to the more extensive literature in classification problems. (R3.1)} Finally, the different loss functions used for Group DRO compared to reweighing and oversampling may have resulted in different regularisation for the models so future work could investigate the use of a single loss function. 

\section{Conclusions}

This paper has reported the most comprehensive investigation of bias mitigation in AI-based CMR segmentation to date.
We have shown that the fairness-accuracy trade-off can be avoided by using a bias understanding-informed approach to mitigation, rather than using a generic mitigation algorithm. Therefore, this represents an important finding that should motivate further investigations into such bias understanding-informed approaches to mitigation in other applications.





\acks{Engineering \& Physical Sciences Research Council Doctoral Training Partnership (EPSRC DTP) grant EP/T517963/1. This research has been conducted using the UK Biobank Resource under Application Number 17806.}

%
\ethics{The work follows appropriate ethical standards in conducting research and writing the manuscript, following all applicable laws and regulations regarding treatment of animals and human subjects}

\coi{E.P-A. is an employee of Heartflow Inc. All work presented is independent of her role at Heartfow. P-G.M. worked as a consultant for Perspectum Diagnostics Ltd until 2023. The remaining authors declare no conflict of interest.}

\data{The UK Biobank dataset is publicly available for approved research projects. Requests to access the dataset should be directed to \url{https://www.ukbiobank.ac.uk/}. The clinical dataset cannot be publicly shared due to limitations of the ethical approval.}

\bibliography{references}

@article{Seyyed-Kalantari2021,
    title = {{CheXclusion: Fairness gaps in deep chest X-ray classifiers}},
    year = {2021},
    journal = {Pacific Symposium on Biocomputing. Pacific Symposium on Biocomputing},
    author = {Seyyed-Kalantari, Laleh and Liu, Guanxiong and McDermott, Matthew and Chen, Irene Y. and Ghassemi, Marzyeh},
    pages = {232--243},
    volume = {26},
    doi = {10.1142/9789811232701{\_}0022},
    issn = {23356936},
    pmid = {33691020},
    arxivId = {2003.00827}
}

@inproceedings{Puyol-Anton2021FairnessSegmentation,
    title = {{Fairness in Cardiac MR Image Analysis: An Investigation of Bias Due to Data Imbalance in Deep Learning Based Segmentation}},
    year = {2021},
    booktitle = {Medical Image Computing and Computer Assisted Intervention – MICCAI 2021},
    author = {Puyol-Ant{\'{o}}n, Esther and Ruijsink, Bram and Piechnik, Stefan K. and Neubauer, Stefan and Petersen, Steffen E. and Razavi, Reza and King, Andrew P.},
    pages = {413--423},
    volume = {12903 LNCS},
    publisher = {Springer International Publishing},
    isbn = {9783030871987},
    doi = {10.1007/978-3-030-87199-4{\_}39},
    issn = {16113349},
    arxivId = {2106.12387}
}

@article{Larrazabal2020,
    title = {{Gender imbalance in medical imaging datasets produces biased classifiers for computer-aided diagnosis}},
    year = {2020},
    journal = {Proceedings of the National Academy of Sciences of the United States of America},
    author = {Larrazabal, Agostina J. and Nieto, Nicolás and Peterson, Victoria and Milone, Diego H. and Ferrante, Enzo},
    number = {23},
    pages = {12592--12594},
    volume = {117},
    isbn = {1919012117},
    doi = {10.1073/pnas.1919012117},
    issn = {10916490},
    pmid = {32457147}
}

@article{Madras2018,
    title = {{Learning adversarially fair and transferable representations}},
    year = {2018},
    journal = {35th International Conference on Machine Learning, ICML 2018},
    author = {Madras, David and Creager, Elliot and Pitassi, Toniann and Zemel, Richards},
    pages = {5423--5434},
    volume = {8},
    isbn = {9781510867963},
    arxivId = {1802.06309}
}

@article{Zhang2018,
    title = {{Mitigating Unwanted Biases with Adversarial Learning}},
    year = {2018},
    journal = {AIES 2018 - Proceedings of the 2018 AAAI/ACM Conference on AI, Ethics, and Society},
    author = {Zhang, Brian Hu and Lemoine, Blake and Mitchell, Margaret},
    pages = {335--340},
    isbn = {9781450360128},
    doi = {10.1145/3278721.3278779},
    issn = {23318422},
    arxivId = {1801.07593}
}

@article{Isensee2020,
    title = {{nnU-Net: a self-configuring method for deep learning-based biomedical image segmentation}},
    year = {2020},
    journal = {Nature Methods 2020 18:2},
    author = {Isensee, Fabian and Jaeger, Paul F. and Kohl, Simon A.A. and Petersen, Jens and Maier-Hein, Klaus H.},
    number = {2},
    month = {12},
    pages = {203--211},
    volume = {18},
    publisher = {Nature Publishing Group},
    doi = {10.1038/s41592-020-01008-z},
    issn = {1548-7105},
    pmid = {33288961}
}

@book{Abbasi-Sureshjani2020,
    title = {{Risk of Training Diagnostic Algorithms on Data with Demographic Bias}},
    year = {2020},
    booktitle = {Lecture Notes in Computer Science (including subseries Lecture Notes in Artificial Intelligence and Lecture Notes in Bioinformatics)},
    author = {Abbasi-Sureshjani, Samaneh and Raumanns, Ralf and Michels, Britt E.J. and Schouten, Gerard and Cheplygina, Veronika},
    pages = {183--192},
    volume = {12446 LNCS},
    publisher = {Springer International Publishing},
    url = {http://dx.doi.org/10.1007/978-3-030-61166-8_20},
    isbn = {9783030611651},
    doi = {10.1007/978-3-030-61166-8{\_}20},
    issn = {16113349},
    arxivId = {2005.10050}
}

@article{Peterson2016,
    title = {{UK Biobank's cardiovascular magnetic resonance protocol}},
    year = {2016},
    journal = {Journal of Cardiovascular Magnetic Resonance},
    author = {Petersen, Steffen E. and Matthews, Paul M. and Francis, Jane M. and Robson, Matthew D. and Zemrak, Filip and Boubertakh, Redha and Young, Alistair A. and Hudson, Sarah and Weale, Peter and Garratt, Steve and Collins, Rory and Piechnik, Stefan and Neubauer, Stefan},
    number = {1},
    month = {2},
    pages = {1--7},
    volume = {18},
    publisher = {BioMed Central Ltd.},
    doi = {10.1186/s12968-016-0227-4},
    issn = {1532429X},
    pmid = {26830817}
}

@article{Seyyed-Kalantari2021a,
    title = {{Underdiagnosis bias of artificial intelligence algorithms applied to chest radiographs in under-served patient populations}},
    year = {2021},
    journal = {Nature Medicine},
    author = {Seyyed-Kalantari, Laleh and Zhang, Haoran and McDermott, Matthew B. A. and Chen, Irene Y. and Ghassemi, Marzyeh},
    number = {12},
    pages = {2176--2182},
    volume = {27},
    publisher = {Springer US},
    doi = {10.1038/s41591-021-01595-0},
    issn = {1078-8956}
}

@article{Balducci2022AFibromyalgia,
    title = {{A behavioral and brain imaging dataset with focus on emotion regulation of women with fibromyalgia}},
    year = {2022},
    journal = {Scientific Data volume},
    author = {Balducci, Thania and Rasgado-Toledo, Jalil and Valencia, Alely and Van Tol, Marie-José and Aleman, André and Garza-Villarreal, Eduardo A},
    number = {581},
    volume = {9},
    url = {www.nature.com/scientificdata},
    doi = {10.1038/s41597-022-01677-9}
}

@article{Ioannou2022ASegmentation,
    title = {{A Study of Demographic Bias in CNN-Based Brain MR Segmentation}},
    year = {2022},
    journal = {Lecture Notes in Computer Science (including subseries Lecture Notes in Artificial Intelligence and Lecture Notes in Bioinformatics)},
    author = {Ioannou, Stefanos and Chockler, Hana and Hammers, Alexander and King, Andrew P.},
    pages = {13--22},
    volume = {13596 LNCS},
    publisher = {Springer, Cham},
    url = {https://link.springer.com/chapter/10.1007/978-3-031-17899-3_2},
    isbn = {978-3-031-17899-3},
    doi = {10.1007/978-3-031-17899-3{\_}2},
    issn = {1611-3349},
    arxivId = {2208.06613}
}

@article{Mehrabi2019ALearning,
    title = {{A Survey on Bias and Fairness in Machine Learning}},
    year = {2019},
    journal = {ACM Computing Surveys},
    author = {Mehrabi, Ninareh and Morstatter, Fred and Saxena, Nripsuta and Lerman, Kristina and Galstyan, Aram},
    number = {6},
    month = {8},
    volume = {54},
    publisher = {Association for Computing Machinery},
    url = {https://arxiv.org/abs/1908.09635v3},
    doi = {10.1145/3457607},
    issn = {15577341},
    arxivId = {1908.09635}
}

@inproceedings{Lee2022ASegmentation,
    title = {{A Systematic Study of Race and Sex Bias in CNN-Based Cardiac MR Segmentation}},
    year = {2022},
    booktitle = {Lecture Notes in Computer Science (including subseries Lecture Notes in Artificial Intelligence and Lecture Notes in Bioinformatics)},
    author = {Lee, Tiarna and Puyol-Ant{\'{o}}n, Esther and Ruijsink, Bram and Shi, Miaojing and King, Andrew P.},
    pages = {233--244},
    volume = {13593 LNCS},
    publisher = {Springer Science and Business Media Deutschland GmbH},
    isbn = {9783031234422},
    doi = {10.1007/978-3-031-23443-9{\_}22},
    issn = {16113349},
    arxivId = {2209.01627}
}

@article{Glocker2023AlgorithmicModels,
    title = {{Algorithmic encoding of protected characteristics in chest X-ray disease detection models}},
    year = {2023},
    journal = {EBioMedicine},
    author = {Glocker, Ben and Jones, Charles and Bernhardt, Mélanie and Winzeck, Stefan},
    month = {3},
    volume = {89},
    publisher = {EBioMedicine},
    url = {https://pubmed.ncbi.nlm.nih.gov/36791660/},
    doi = {10.1016/J.EBIOM.2023.104467},
    issn = {2352-3964},
    pmid = {36791660}
}

@article{Lee2025AnSegmentation,
    title = {{An investigation into the causes of race bias in AI-based cine CMR segmentation}},
    year = {2025},
    journal = {European Heart Journal - Digital Health},
    author = {Lee, Tiarna and Puyol-Ant{\'{o}}n, Esther and Ruijsink, Bram and Roujol, Sebastien and Barfoot, Theodore and Ogbomo-Harmitt, Shaheim and Shi, Miaojing and King, Andrew},
    month = {2},
    url = {https://dx.doi.org/10.1093/ehjdh/ztaf008},
    doi = {10.1093/EHJDH/ZTAF008},
    issn = {2634-3916}
}

@article{Lee2023AnSegmentation,
    title = {{An Investigation into the Impact of Deep Learning Model Choice on Sex and Race Bias in Cardiac MR Segmentation}},
    year = {2023},
    journal = {Lecture Notes in Computer Science (including subseries Lecture Notes in Artificial Intelligence and Lecture Notes in Bioinformatics)},
    author = {Lee, Tiarna and Puyol-Ant{\'{o}}n, Esther and Ruijsink, Bram and Aitcheson, Keana and Shi, Miaojing and King, Andrew P.},
    pages = {215--224},
    volume = {14242 LNCS},
    publisher = {Springer Science and Business Media Deutschland GmbH},
    url = {https://link.springer.com/chapter/10.1007/978-3-031-45249-9_21},
    isbn = {9783031452482},
    doi = {10.1007/978-3-031-45249-9{\_}21/FIGURES/2},
    issn = {16113349}
}

@article{Wang2023BiasStudies,
    title = {{Bias in machine learning models can be significantly mitigated by careful training: Evidence from neuroimaging studies}},
    year = {2023},
    journal = {Proceedings of the National Academy of Sciences of the United States of America},
    author = {Wang, Rongguang and Chaudhari, Pratik and Davatzikos, Christos},
    number = {6},
    month = {2},
    pages = {e2211613120},
    volume = {120},
    publisher = {National Academy of Sciences},
    url = {https://www.pnas.org/doi/abs/10.1073/pnas.2211613120},
    doi = {10.1073/PNAS.2211613120/SUPPL{\_}FILE/PNAS.2211613120.SAPP.PDF},
    issn = {10916490},
    pmid = {36716365}
}

@article{Lohia2019BiasFairness,
    title = {{Bias Mitigation Post-processing for Individual and Group Fairness}},
    year = {2019},
    journal = {ICASSP, IEEE International Conference on Acoustics, Speech and Signal Processing - Proceedings},
    author = {Lohia, Pranay K. and Natesan Ramamurthy, Karthikeyan and Bhide, Manish and Saha, Diptikalyan and Varshney, Kush R. and Puri, Ruchir},
    month = {5},
    pages = {2847--2851},
    volume = {2019-May},
    publisher = {Institute of Electrical and Electronics Engineers Inc.},
    isbn = {9781479981311},
    doi = {10.1109/ICASSP.2019.8682620},
    issn = {15206149},
    arxivId = {1812.06135}
}

@article{Kamiran2012DataDiscrimination,
    title = {{Data preprocessing techniques for classification without discrimination}},
    year = {2012},
    journal = {Knowledge and Information Systems},
    author = {Kamiran, Faisal and Calders, Toon},
    number = {1},
    month = {12},
    pages = {1--33},
    volume = {33},
    publisher = {Springer London},
    url = {https://link.springer.com/article/10.1007/s10115-011-0463-8},
    doi = {10.1007/S10115-011-0463-8/METRICS},
    issn = {02193116}
}

@article{Kamiran2012DecisionClassification,
    title = {{Decision theory for discrimination-aware classification}},
    year = {2012},
    journal = {Proceedings - IEEE International Conference on Data Mining, ICDM},
    author = {Kamiran, Faisal and Karim, Asim and Zhang, Xiangliang},
    pages = {924--929},
    isbn = {9780769549057},
    doi = {10.1109/ICDM.2012.45},
    issn = {15504786}
}

@article{Schrouff2022DiagnosingSettings,
    title = {{Diagnosing failures of fairness transfer across distribution shift in real-world medical settings}},
    year = {2022},
    journal = {Advances in Neural Information Processing Systems},
    author = {Schrouff, Jessica and Harris, Natalie and Research, Google and Koyejo, Oluwasanmi and Alabdulmohsin, Ibrahim and Schnider, Eva and Opsahl-Ong, Krista and Brown, Alex and Roy, Subhrajit and Mincu, Diana and Chen, Christina and Dieng, Awa and Liu, Yuan and Natarajan, Vivek and Karthikesalingam, Alan and Heller, Katherine and Chiappa, Silvia and Alexander D', Deepmind and Research, Amour Google},
    month = {12},
    pages = {19304--19318},
    volume = {35}
}

@article{Daneshjou2022DisparitiesSet,
    title = {{Disparities in dermatology AI performance on a diverse, curated clinical image set}},
    year = {2022},
    journal = {Science Advances},
    author = {Daneshjou, Roxana and Vodrahalli, Kailas and Novoa, Roberto A. and Jenkins, Melissa and Liang, Weixin and Rotemberg, Veronica and Ko, Justin and Swetter, Susan M. and Bailey, Elizabeth E. and Gevaert, Olivier and Mukherjee, Pritam and Phung, Michelle and Yekrang, Kiana and Fong, Bradley and Sahasrabudhe, Rachna and Allerup, Johan A.C. and Okata-Karigane, Utako and Zou, James and Chiou, Albert S.},
    number = {32},
    month = {8},
    pages = {6147},
    volume = {8},
    publisher = {American Association for the Advancement of Science},
    url = {https://www.science.org/doi/10.1126/sciadv.abq6147},
    doi = {10.1126/SCIADV.ABQ6147/SUPPL{\_}FILE/SCIADV.ABQ6147{\_}SM.PDF},
    issn = {23752548},
    pmid = {35960806},
    arxivId = {2203.08807}
}

@inproceedings{Sagawa2020DISTRIBUTIONALLYGENERALIZATION,
    title = {{DISTRIBUTIONALLY ROBUST NEURAL NETWORKS FOR GROUP SHIFTS: ON THE IMPORTANCE OF REGULARIZATION FOR WORST-CASE GENERALIZATION}},
    year = {2020},
    booktitle = {8th International Conference on Learning Representations, ICLR 2020},
    author = {Sagawa, Shiori and Koh, Pang Wei and Hashimoto, Tatsunori B. and Liang, Percy}
}

@article{Lee2025DoesSegmentation,
    title = {{Does a Rising Tide Lift All Boats? Bias Mitigation for AI-based CMR Segmentation}},
    year = {2025},
    journal = {arXiv},
    author = {Lee, Tiarna and Puyol-Ant{\'{o}}n, Esther and Ruijsink, Bram and Shi, Miaojing and King, Andrew P.},
    month = {3},
    url = {https://arxiv.org/abs/2503.17089v1},
    arxivId = {2503.17089}
}

@article{Georgiopoulos2024EthnicityResonance,
    title = {{Ethnicity differences in geometric remodelling and myocardial composition in hypertension unveiled by cardiovascular magnetic resonance}},
    year = {2024},
    journal = {European Heart Journal - Cardiovascular Imaging},
    author = {Georgiopoulos, Georgios and Faconti, Luca and Mohamed, Aqeel T. and Figliozzi, Stefano and Asher, Clint and Keehn, Louise and McNally, Ryan and Alfakih, Khaled and Vennin, Samuel and Chiribiri, Amedeo and Lamata, Pablo and Chowienczyk, Philip and Masci, Pier Giorgio},
    number = {7},
    month = {6},
    pages = {901--911},
    volume = {25},
    publisher = {Oxford Academic},
    url = {https://dx.doi.org/10.1093/ehjci/jeae097},
    doi = {10.1093/EHJCI/JEAE097},
    issn = {2047-2404},
    pmid = {38597630}
}

@article{Adewole2024ExpandingDataset,
    title = {{Expanding the Brain Tumor Segmentation (BraTS) data to include African Populations (BraTS-Africa) (version 1) [Dataset]}},
    year = {2024},
    journal = {The Cancer Imaging Archive},
    author = {Adewole, M. and Rudie, J.D. and Gbadamosi, A. and Zhang, D. and Raymond, C. and Ajigbotoshso, J. and Toyobo, O. and Aguh, K. and Omidiji, O. and {Akinola R.} and Suwaid, M.A. and Emegoakor, A. and Ojo, N. and Kalaiwo, C. and {Babatunde G.} and Ogunleye, A. and Gbadamosi, Y. and Iorpagher, K. and {Onuwaje M.} and {Betiku B.} and Saluja, R. and Menze, B. and Baid, U. and Bakas, S. and Dako, F. and {Fatade A.} and Anazodo, U.C.},
    url = {https://www.cancerimagingarchive.net/collection/brats-africa/},
    arxivId = {DOI: 10.7937/v8h6-8×67}
}

@article{Siddiqui2024FairHealthcare,
    title = {{Fair AI-powered orthopedic image segmentation: addressing bias and promoting equitable healthcare}},
    year = {2024},
    journal = {Scientific Reports |},
    author = {Siddiqui, Ismaeel A and Littlefield, Nickolas and Carlson, Luke A and Gong, Matthew and Chhabra, Avani and Menezes, Zoe and Mastorakos, George M and Mehul Thakar, Sakshi and Abedian, Mehrnaz and Lohse, Ines and Weiss, Kurt R and Plate, Johannes F and Moradi, Hamidreza and Amirian, Soheyla and Tafti, Ahmad P},
    pages = {16105},
    volume = {14},
    url = {https://doi.org/10.1038/s41598-024-66873-6},
    isbn = {0123456789},
    doi = {10.1038/s41598-024-66873-6}
}

@article{Puyol-Anton2022FairnessSegmentation,
    title = {{Fairness in Cardiac Magnetic Resonance Imaging: Assessing Sex and Racial Bias in Deep Learning-Based Segmentation}},
    year = {2022},
    journal = {Frontiers in Cardiovascular Medicine},
    author = {Puyol-Ant{\'{o}}n, Esther and Ruijsink, Bram and Mariscal Harana, Jorge and Piechnik, Stefan K. and Neubauer, Stefan and Petersen, Steffen E. and Razavi, Reza and Chowienczyk, Phil and King, Andrew P.},
    month = {4},
    pages = {664},
    volume = {0},
    publisher = {Frontiers},
    doi = {10.3389/FCVM.2022.859310},
    issn = {2297-055X}
}

@article{Tian2023FairSeg:Scaling,
    title = {{FairSeg: A Large-Scale Medical Image Segmentation Dataset for Fairness Learning Using Segment Anything Model with Fair Error-Bound Scaling}},
    year = {2023},
    journal = {arXiv},
    author = {Tian, Yu and Shi, Min and Luo, Yan and Kouhana, Ava and Elze, Tobias and Wang, Mengyu},
    month = {11},
    url = {https://arxiv.org/abs/2311.02189v5},
    arxivId = {2311.02189}
}

@article{Petersen2022FeatureDetection,
    title = {{Feature Robustness and Sex Differences in Medical Imaging: A Case Study in MRI-Based Alzheimer’s Disease Detection}},
    year = {2022},
    journal = {Lecture Notes in Computer Science (including subseries Lecture Notes in Artificial Intelligence and Lecture Notes in Bioinformatics)},
    author = {Petersen, Eike and Feragen, Aasa and da Costa Zemsch, Maria Luise and Henriksen, Anders and Wiese Christensen, Oskar Eiler and Ganz, Melanie},
    pages = {88--98},
    volume = {13431 LNCS},
    publisher = {Springer Science and Business Media Deutschland GmbH},
    url = {https://link.springer.com/chapter/10.1007/978-3-031-16431-6_9},
    isbn = {9783031164309},
    doi = {10.1007/978-3-031-16431-6{\_}9/FIGURES/3},
    issn = {16113349},
    arxivId = {2204.01737}
}

@article{Klingenberg2023HigherDetection,
    title = {{Higher performance for women than men in MRI-based Alzheimer’s disease detection}},
    year = {2023},
    journal = {Alzheimer's Research and Therapy},
    author = {Klingenberg, Malte and Stark, Didem and Eitel, Fabian and Budding, Céline and Habes, Mohamad and Ritter, Kerstin},
    number = {1},
    month = {12},
    pages = {1--13},
    volume = {15},
    publisher = {BioMed Central Ltd},
    url = {https://alzres.biomedcentral.com/articles/10.1186/s13195-023-01225-6 http://creativecommons.org/publicdomain/zero/1.0/},
    doi = {10.1186/S13195-023-01225-6/FIGURES/6},
    issn = {17589193},
    pmid = {37081528}
}

@article{Zhang2022ImprovingClassifiers,
    title = {{Improving the Fairness of Chest X-ray Classifiers}},
    year = {2022},
    journal = {arXiv},
    author = {Zhang, Haoran and Dullerud, Natalie and Roth, Karsten and Oakden-Rayner, Lauren and Pfohl, Stephen Robert and Ghassemi, Marzyeh},
    month = {3},
    pages = {2022},
    url = {https://arxiv.org/abs/2203.12609v1},
    doi = {10.48550/arxiv.2203.12609},
    arxivId = {2203.12609}
}

@article{He2017MaskR-CNN,
    title = {{Mask R-CNN}},
    year = {2017},
    journal = {IEEE Transactions on Pattern Analysis and Machine Intelligence},
    author = {He, Kaiming and Gkioxari, Georgia and Doll{\'{a}}r, Piotr and Girshick, Ross},
    number = {2},
    month = {3},
    pages = {386--397},
    volume = {42},
    publisher = {IEEE Computer Society},
    url = {https://arxiv.org/abs/1703.06870v3},
    doi = {10.1109/TPAMI.2018.2844175},
    issn = {19393539},
    pmid = {29994331},
    arxivId = {1703.06870}
}

@article{Zong2022MEDFAIR:Imaging,
    title = {{MEDFAIR: Benchmarking Fairness for Medical Imaging}},
    year = {2022},
    journal = {arXiv},
    author = {Zong, Yongshuo and Yang, Yongxin and Hospedales, Timothy},
    month = {10},
    url = {https://arxiv.org/abs/2210.01725v1},
    doi = {10.48550/arxiv.2210.01725},
    arxivId = {2210.01725}
}

@article{Martinez2020MinimaxPerspective,
    title = {{Minimax Pareto Fairness: A Multi Objective Perspective}},
    year = {2020},
    journal = {Proceedings of machine learning research},
    author = {Martinez, Natalia and Bertran, Martin and Sapiro, Guillermo},
    pages = {6755},
    volume = {119},
    publisher = {International Machine Learning Society (IMLS)},
    url = {https://pmc.ncbi.nlm.nih.gov/articles/PMC7912461/},
    isbn = {9781713821120},
    issn = {2640-3498},
    pmid = {33644764},
    arxivId = {2011.01821}
}

@article{Gottfrois2024PASSIONAfrica,
    title = {{PASSION for Dermatology: Bridging the Diversity Gap with Pigmented Skin Images from Sub-Saharan Africa}},
    year = {2024},
    journal = {Medical Image Computing and Computer Assisted Intervention – MICCAI 2024},
    author = {Gottfrois, Philippe and Gr{\"{o}}ger, Fabian and Andriambololoniaina, Faly Herizo and Amruthalingam, Ludovic and Gonzalez-Jimenez, Alvaro and Hsu, Christophe and Kessy, Agnes and Lionetti, Simone and Mavura, Daudi and Ng’ambi, Wingston and Ngongonda, Dingase Faith and Pouly, Marc and Rakotoarisaona, Mendrika Fifaliana and Rapelanoro Rabenja, Fahafahantsoa and Traor{\'{e}}, Ibrahima and Navarini, Alexander A.},
    pages = {703--712},
    publisher = {Springer, Cham},
    url = {https://link.springer.com/chapter/10.1007/978-3-031-72384-1_66},
    isbn = {978-3-031-72384-1},
    doi = {10.1007/978-3-031-72384-1{\_}66},
    issn = {1611-3349}
}

@article{Barbieri2022ReimaginingData,
    title = {{Reimagining leprosy elimination with AI analysis of a combination of skin lesion images with demographic and clinical data}},
    year = {2022},
    journal = {Lancet regional health. Americas},
    author = {Barbieri, Raquel R. and Xu, Yixi and Setian, Lucy and Souza-Santos, Paulo Thiago and Trivedi, Anusua and Cristofono, Jim and Bhering, Ricardo and White, Kevin and Sales, Anna M. and Miller, Geralyn and Nery, José Augusto C. and Sharman, Michael and Bumann, Richard and Zhang, Shun and Goldust, Mohamad and Sarno, Euzenir N. and Mirza, Fareed and Cavaliero, Arielle and Timmer, Sander and Bonfiglioli, Elena and Smith, Cairns and Scollard, David and Navarini, Alexander A. and Aerts, Ann and Ferres, Juan Lavista and Moraes, Milton O.},
    month = {5},
    volume = {9},
    publisher = {Lancet Reg Health Am},
    url = {https://pubmed.ncbi.nlm.nih.gov/36776278/},
    doi = {10.1016/J.LANA.2022.100192},
    issn = {2667-193X},
    pmid = {36776278}
}

@article{Olesen2025SlicingMethods,
    title = {{Slicing Through Bias: Explaining Performance Gaps in Medical Image Analysis Using Slice Discovery Methods}},
    year = {2025},
    journal = {Lecture Notes in Computer Science (including subseries Lecture Notes in Artificial Intelligence and Lecture Notes in Bioinformatics) },
    author = {Olesen, Vincent and Weng, Nina and Feragen, Aasa and Petersen, Eike},
    pages = {3--13},
    volume = {15198 LNCS},
    publisher = {Springer, Cham},
    url = {https://link.springer.com/chapter/10.1007/978-3-031-72787-0_1},
    isbn = {978-3-031-72787-0},
    doi = {10.1007/978-3-031-72787-0{\_}1},
    issn = {1611-3349}
}

@article{Yang2024TheGeneralization,
    title = {{The limits of fair medical imaging AI in real-world generalization}},
    year = {2024},
    journal = {Nature Medicine 2024 30:10},
    author = {Yang, Yuzhe and Zhang, Haoran and Gichoya, Judy W. and Katabi, Dina and Ghassemi, Marzyeh},
    number = {10},
    month = {6},
    pages = {2838--2848},
    volume = {30},
    publisher = {Nature Publishing Group},
    url = {https://www.nature.com/articles/s41591-024-03113-4},
    doi = {10.1038/s41591-024-03113-4},
    issn = {1546-170X}
}

@article{Li2024TheSurvey,
    title = {{The Triangular Trade-off between Robustness, Accuracy and Fairness in Deep Neural Networks: A Survey}},
    year = {2024},
    journal = {ACM Computing Surveys},
    author = {Li, Jingyang and Li, Guoqiang},
    doi = {10.1145/3645088},
    issn = {0360-0300}
}

@article{Bencevic2024UnderstandingSegmentation,
    title = {{Understanding skin color bias in deep learning-based skin lesion segmentation}},
    year = {2024},
    journal = {Computer Methods and Programs in Biomedicine},
    author = {Ben{\v{c}}evi{\'{c}}, Marin and Habijan, Marija and Gali{\'{c}}, Irena and Babin, Danilo and Pi{\v{z}}urica, Aleksandra},
    month = {3},
    pages = {108044},
    volume = {245},
    publisher = {Elsevier},
    doi = {10.1016/J.CMPB.2024.108044},
    issn = {0169-2607},
    pmid = {38290289}
}





\clearpage


\setcounter{figure}{0}
\setcounter{table}{0}

\makeatletter
\renewcommand{\thefigure}{S\arabic{figure}}
\renewcommand{\thetable}{S\arabic{table}}

\renewcommand{\theHfigure}{S\arabic{figure}}
\renewcommand{\theHtable}{S\arabic{table}}
\makeatother

\ifcsname crefname\endcsname
  \crefname{figure}{Figure}{Figures}
  \crefname{table}{Table}{Tables}
  \Crefformat{figure}{Figure~#2S#1#3}   
  \Crefformat{table}{Table~#2S#1#3}
  \crefformat{figure}{Figure~#2S#1#3}
  \crefformat{table}{Table~#2S#1#3}
\fi

\section*{Supplementary Material}
\addcontentsline{toc}{section}{Supplementary Material}
\label{sec:supplementary}

\begin{table}[!ht]
\centering
 \textbf{\large Metrics for regions of the heart using the external validation set}\par\vspace{0.3em}
\caption{Metrics for LVBP segmentation based on inference performed on external validation set using different models trained on UK Biobank data. The p-values were computed between White and Black subjects based on a two-sided Mann Whitney U test on the DSC scores. * p$<$ 0.05. The best median DSC score for Black and White subjects is shown in bold.}
\begin{adjustbox}{width=\columnwidth,center}
\begin{tabular}{@{}|c|cc|cc|cc|cc|@{}|}
\hline
&  \multicolumn{2}{c|}{\begin{tabular}[c]{@{}c@{}}Baseline \\ p = 0.11 \end{tabular}} & \multicolumn{2}{|c|}{\begin{tabular}[c]{@{}c@{}}Oversampling \\ p = 0.46 \end{tabular}}  \\ \hline
&White  &Black  &White  &Black  \\\hline

Median DSC & \textbf{0.940}& 0.931 & 0.936& \textbf{0.937} \\
IQR DSC & 0.041& 0.035 & 0.038& 0.032 \\
Median HD (mm) & 3.929 & 3.416 & 3.705 & 3.518 \\
IQR HD (mm) & 1.622 & 1.498 & 1.296 & 2.186 \\
SER     &\multicolumn{2}{c|}{1.146} &\multicolumn{2}{c|}{1.007} \\
Fairness gap    &\multicolumn{2}{c|}{0.009} &\multicolumn{2}{c|}{-0.0004} \\\hline
\end{tabular}
\label{tab:LVBP_results}
\end{adjustbox}
\end{table}

\begin{table}[!ht]
\centering
\caption{Metrics for LVM segmentation based on inference performed on the external validation set using different models trained on UK Biobank data. The p-values were computed between White and Black subjects based on a two-sided Mann Whitney U test on the DSC scores. * p$<$ 0.05. The best median DSC score for Black and White subjects is shown in bold.}
\begin{adjustbox}{width=\columnwidth,center}
\begin{tabular}{@{}|c|cc|cc|cc|cc|@{}|}
\hline
&  \multicolumn{2}{c|}{\begin{tabular}[c]{@{}c@{}}Baseline \\ p = 0.84 \end{tabular}} & \multicolumn{2}{|c|}{\begin{tabular}[c]{@{}c@{}}Oversampling \\ p = 0.57 \end{tabular}}  \\ \hline
&White  &Black  &White  &Black \\\hline

Median DSC &  \textbf{0.817 }& \textbf{0.803}& 0.808 & 0.802 \\
IQR DSC & 0.046& 0.075 & 0.053& 0.06\\
Median HD (mm) & 5.026 & 5.141 & 5.096 & 5.318 \\
IQR HD (mm) & 1.583 & 1.941 & 1.995 & 2.042\\
SER     &\multicolumn{2}{c|}{1.077} &\multicolumn{2}{c|}{1.034} \\
Fairness gap    &\multicolumn{2}{c|}{0.014} &\multicolumn{2}{c|}{0.006} \\\hline
\end{tabular}
\label{tab: LVM results}
\end{adjustbox}
\end{table}

\begin{table}[!ht]
\centering
\caption{Metrics for RVBP segmentation based on inference performed on the external validation set using different models trained on UK Biobank data. The p-values were computed between White and Black subjects based on a two-sided Mann Whitney U test on the DSC scores. * p$<$ 0.05. The best median DSC score for Black and White subjects is shown in bold.}
\begin{adjustbox}{width=\columnwidth,center}
\begin{tabular}{@{}|c|cc|cc|cc|cc|@{}|}
\hline
&  \multicolumn{2}{c|}{\begin{tabular}[c]{@{}c@{}}Baseline \\ p = 0.43 \end{tabular}} & \multicolumn{2}{|c|}{\begin{tabular}[c]{@{}c@{}}Oversampling \\ p = 0.27 \end{tabular}}  \\ \hline
&White  &Black  &White  &Black  \\\hline

Median DSC & 0.883& \textbf{0.882} & \textbf{0.885}& 0.868 \\
IQR DSC & 0.057& 0.048 & 0.070& 0.071  \\
Median HD (mm) & 8.902 & 7.783 & 8.869 & 10.181  \\
IQR HD (mm) & 3.208 & 3.096 & 3.962 & 4.401\\
SER     &\multicolumn{2}{c|}{1.004} &\multicolumn{2}{c|}{1.148} \\
Fairness gap    &\multicolumn{2}{c|}{0.0004} &\multicolumn{2}{c|}{0.017}\\\hline
\end{tabular}
\label{tab:RVBP_results}
\end{adjustbox}
\end{table}

\begin{table}[!ht]
 \textbf{\large Difference in the size of the bounding box between images}\par\vspace{0.3em}
\caption{Difference in the size of the bounding box between images cropped using the ground truth and cascaded approach for the internal set.}
\label{tab:differences_in_bounding_box}
\begin{tabular}{|c|c|c|}
\hline
Experiment   & Error $x$ (\%) & Error $y$ (\%) \\ \hline
Baseline     & 3.70           & 6.57           \\\hline
Oversampling & 0.0            & 7.30           \\\hline
Reweighing   & 1.48           & 8.76           \\\hline
Group DRO    & 0.0            & 0.730        \\ \hline
\end{tabular}
\end{table}

\begin{table}[ht]
\centering
 \textbf{\large Metrics for regions of the heart for the external validation set using the cascaded cropping method}\par\vspace{0.3em}
\caption{Metrics for LVBP segmentation based on inference performed on the external validation set using the cascaded approach to crop the images. The p-values were computed between White and Black subjects based on a two-sided Mann Whitney U test on the DSC scores. * p$<$ 0.05. The best median DSC score for Black and White subjects is shown in bold.}
\begin{adjustbox}{width=\columnwidth,center}
\begin{tabular}{@{}|c|cc|cc|cc|cc|cc|cc|}
\hline & \multicolumn{2}{c|}{\begin{tabular}[c]{@{}c@{}}Cropped baseline \\ p = 0.088 \end{tabular}} & \multicolumn{2}{c|}{\begin{tabular}[c]{@{}c@{}}Cropped oversampling \\ p = 0.76 \end{tabular}}
& \multicolumn{2}{c}{\begin{tabular}[c]{@{}c@{}}Cropped baseline \\ cascaded model \\ * \end{tabular}}& \multicolumn{2}{|c|}{\begin{tabular}[c]{@{}c@{}}Cropped oversampling \\ cascaded model \\ p = 0.72\end{tabular}}\\ \hline
&White  &Black  &White  &Black &White  &Black  &White  &Black\\\hline

Median DSC & \textbf{0.962}& \textbf{0.949}  & 0.958 & 0.943 & 0.959 & 0.943 & 0.959 & 0.946\\
IQR DSC & 0.030& 0.031 & 0.024 & 0.027  & 0.031 & 0.029 & 0.036 & 0.024 \\
Median HD (mm) & 2.887 & 2.488 & 3.043 & 2.870 & 3.132 & 2.839 & 2.842 & 2.493\\
IQR HD (mm) & 0.984 & 0.854 & 1.113 & 1.697 & 1.119 & 1.049 & 1.352 & 2.646 \\
SER    &\multicolumn{2}{c|}{1.309} &\multicolumn{2}{c|}{1.364} &\multicolumn{2}{c|}{1.398} &\multicolumn{2}{c|}{1.316} \\
Fairness gap &\multicolumn{2}{c|}{0.012}  &\multicolumn{2}{c|}{0.015}  &\multicolumn{2}{c|}{0.016} &\multicolumn{2}{c|}{0.013} \\\hline
\end{tabular}
\label{tab:LVBP_cascaded_results}
\end{adjustbox}
\end{table}

\begin{table}[ht]
\centering
\caption{Metrics for LVM segmentation based on inference performed on external validation set using the cascaded approach to crop the images. The p-values were computed between White and Black subjects based on a two-sided Mann Whitney U test on the DSC scores. * p$<$ 0.05. The best median DSC score for Black and White subjects is shown in bold.}
\begin{adjustbox}{width=\columnwidth,center}
\begin{tabular}{@{}|c|cc|cc|cc|cc|cc|cc|}
\hline & \multicolumn{2}{c|}{\begin{tabular}[c]{@{}c@{}}Cropped baseline \\ p = 0.39 \end{tabular}} & \multicolumn{2}{c|}{\begin{tabular}[c]{@{}c@{}}Cropped oversampling \\ p = 0.63 \end{tabular}}
& \multicolumn{2}{c}{\begin{tabular}[c]{@{}c@{}}Cropped baseline \\ cascaded model \\ p = 0.69\end{tabular}}& \multicolumn{2}{|c|}{\begin{tabular}[c]{@{}c@{}}Cropped oversampling \\ cascaded model \\ p = 0.86 \end{tabular}}\\ \hline
&White  &Black  &White  &Black  &White  &Black  &White  &Black\\\hline

Median DSC & \textbf{0.881 }& 0.860& 0.852 & 0.849 & 0.856 & 0.855 & 0.879 & \textbf{0.863}\\
IQR DSC 7& 0.103& 0.093 & 0.096 & 0.083 & 0.093 & 0.106 & 0.071 & 0.066\\
Median HD (mm) & 3.512 & 3.310 & 4.071 & 3.937 & 4.051 & 4.008  & 3.422 & 3.259 \\
IQR HD (mm)  & 1.738 & 1.251 & 2.030 & 2.492 & 2.532 & 2.098  & 2.051 & 3.120 \\
SER    &\multicolumn{2}{c|}{1.175} &\multicolumn{2}{c|}{1.019}  &\multicolumn{2}{c|}{1.009} &\multicolumn{2}{c|}{1.130} \\
Fairness gap   &\multicolumn{2}{c|}{0.021}  &\multicolumn{2}{c|}{0.003} &\multicolumn{2}{c|}{0.001} &\multicolumn{2}{c|}{0.016} \\\hline
\end{tabular}
\label{tab: LVM cascaded results}
\end{adjustbox}
\end{table}

\begin{table}[ht]
\centering
\caption{Metrics for RVBP segmentation based on inference performed on external validation set using the cascaded approach to crop the images. The p-values were computed between White and Black subjects based on a two-sided Mann Whitney U test on the DSC scores. * p$<$ 0.05. The best median DSC score for Black and White subjects is shown in bold.}
\label{tab:RVBP_cascaded_results}

\begin{adjustbox}{width=\columnwidth,center}
\begin{tabular}{@{}|c|cc|cc|cc|cc|cc|cc|}
\hline & \multicolumn{2}{c|}{\begin{tabular}[c]{@{}c@{}}Cropped baseline \\ p = 0.75 \end{tabular}} & \multicolumn{2}{c|}{\begin{tabular}[c]{@{}c@{}}Cropped oversampling \\ p = 0.71 \end{tabular}}
& \multicolumn{2}{c}{\begin{tabular}[c]{@{}c@{}}Cropped baseline \\ cascaded model \\ p =0.84\end{tabular}}& \multicolumn{2}{|c|}{\begin{tabular}[c]{@{}c@{}}Cropped oversampling \\ cascaded model \\ *\end{tabular}}\\ \hline
&White  &Black  &White  &Black &White  &Black  &White  &Black \\\hline

Median DSC & \textbf{0.938}& 0.925& 0.918 & 0.907 & 0.912 & 0.905 & 0.929 & \textbf{0.932} \\
IQR DSC & 0.024& 0.022 & 0.034 & 0.026 & 0.045 & 0.040 & 0.049 & 0.031\\
Median HD (mm) & 5.932 & 5.787  & 6.540 & 6.662 & 7.477 & 7.147 & 4.871 & 6.426\\
IQR HD (mm)  & 2.580 & 3.557 & 2.991 & 2.753 & 4.459 & 5.708 & 2.018 & 4.328\\
SER     &\multicolumn{2}{c|}{1.197} &\multicolumn{2}{c|}{1.136} &\multicolumn{2}{c|}{ 1.131} &\multicolumn{2}{c|}{1.036} \\
Fairness gap    &\multicolumn{2}{c|}{0.012}  &\multicolumn{2}{c|}{0.011} &\multicolumn{2}{c|}{0.011} &\multicolumn{2}{c|}{-0.003} \\\hline
\end{tabular}
\end{adjustbox}
\end{table}

\begin{table}[]
\centering
 \textbf{\large DSC and HD values by sex and age for the cascaded approach}\par\vspace{0.3em}
\caption{DSC and HD values by sex and age for the cascaded approach using oversampling on the external validation set. The p-values were computed based on a two-sided Mann Whitney U test on the DSC scores. * p$<$ 0.05. FG was calculated using $DSC_{male} - DSC_{female}$ and $DSC_{Age < 50 } - DSC_{Age > 50}$}
\label{tab:overall_cascaded_results_sex_and_age}

\begin{adjustbox}{width=\columnwidth,center}
\begin{tabular}{@{}|c|cc|cc|cc|cc|}
\hline
& \multicolumn{2}{c}{\begin{tabular}[c]{@{}c@{}}Cropped baseline \\ cascaded model \\ p = 0.12 \end{tabular}}& \multicolumn{2}{|c|}{\begin{tabular}[c]{@{}c@{}}Cropped baseline \\ cascaded model \\ p = 0.68\end{tabular}}\\ \hline
&Female  &Male  & Age $<$ 50  & Age $>$ 50 \\\hline

n & 28 & 56 & 52 & 32 \\\hline

Median DSC & 0.913 & 0.927 & 0.922 & 0.922\\
IQR DSC  & 0.032 & 0.036  & 0.032 & 0.031 \\
Median HD (mm) & 4.069 & 4.143 & 3.869 & 4.208 \\ 
IQR HD (mm) & 1.927 & 1.402 & 1.558 & 1.738 \\
SER     &\multicolumn{2}{c|}{1.193} &\multicolumn{2}{c|}{1.004} \\
Fairness gap    &\multicolumn{2}{c|}{0.0141} &\multicolumn{2}{c|}{-0.0003} \\\hline
\end{tabular}
\end{adjustbox}
\end{table}

\begin{table*}[!ht]    
\centering
 \textbf{\large Metrics for the bias mitigation methods using White and Asian subjects}\par\vspace{0.3em}
\caption{DSC and HD values for each of the bias mitigation methods tested on the internal validation set. The p-values were computed between White and Asian subjects based on a two-sided Mann Whitney U test  on the DSC scores. * p$<$ 0.05. The best median DSC score for Asian and White subjects is shown in bold.}
\label{tab:bias-mitigation-results-aw}

\begin{adjustbox}{width=\textwidth,center}

\begin{tabular}{|c|c|c|c|c|c|c|c|c|c|c|c|c|}
\hline
& \multicolumn{2}{c}{Baseline *} & \multicolumn{2}{|c|}{\begin{tabular}[c]{@{}c@{}}Oversampling \\ p = 0.17\end{tabular}} & \multicolumn{2}{c}{Reweighing *} & \multicolumn{2}{|c|}{Group DRO *} & \multicolumn{2}{|c|}{\begin{tabular}[c]{@{}c@{}}Cropped baseline \\ cascaded model \\ * \end{tabular}}& \multicolumn{2}{|c|}{\begin{tabular}[c]{@{}c@{}}Cropped oversampling \\ cascaded model \\ * \end{tabular}} \\
\hline
& White  & Asian  & White  & Asian  & White  & Asian  & White  & Asian & White  & Asian  & White  & Asian \\
\hline
Median DSC & 0.901 & 0.851 & 0.8990 & 0.9039 & 0.8971 & 0.8349 &0.8971 & 0.8704 & \textbf{0.9584} & \textbf{0.9362} & 0.9576 & 0.8949 \\

IQR DSC & 0.0452 & 0.0635 & 0.0445 & 0.0678 & 0.0501 & 0.0686 & 0.0471 & 0.0584 & 0.0559 & 0.0502 & 0.0581 & 0.0583 \\

Median HD (mm) & 6.761 & 9.585 & 6.783 & 7.204 & 7.051 & 10.057 & 6.688 & 8.628 & 3.957 & 7.254 & 3.862 & 8.093\\
IQR HD (mm) & 3.690 & 4.935 & 3.715 & 3.662 & 4.140 & 5.076 & 3.744 & 4.310 & 2.470 & 5.115 & 2.321 & 5.491\\

SER     &\multicolumn{2}{c}{1.498}  &\multicolumn{2}{|c|}{1.051}  &\multicolumn{2}{c}{1.604}        &\multicolumn{2}{|c|}{1.260} &\multicolumn{2}{|c|}{1.533} &\multicolumn{2}{|c|}{2.477}\\ 
Fairness gap    &\multicolumn{2}{c}{0.0494}  &\multicolumn{2}{|c|}{-0.0049}  &\multicolumn{2}{c}{0.0621}  &\multicolumn{2}{|c|}{0.0267} &\multicolumn{2}{|c|}{0.0222} &\multicolumn{2}{|c|}{0.0627}\\\hline 
\end{tabular}
\end{adjustbox}
\end{table*}

\begin{figure*}[!hb]
    \centering
     \textbf{\large The effect of changing the number of Black subjects in the training dataset}\par\vspace{0.3em}
    \begin{subfigure}{0.49\textwidth}
        \centering
        \includegraphics[width=\textwidth]{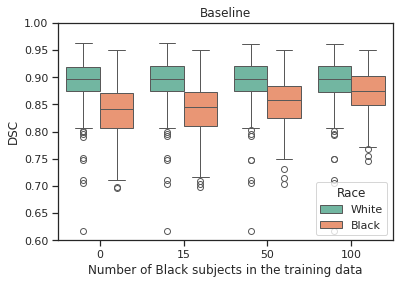}
        \label{fig:No black subjects baseline}
    \end{subfigure}
    \hfill
    \begin{subfigure}{0.49\textwidth}
        \centering
        \includegraphics[width=\textwidth]{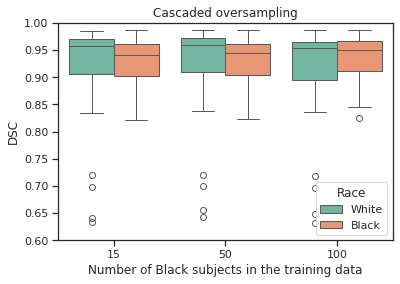}
        \label{fig:No black subjects cascaded}
    \end{subfigure}
    \caption{The effect of changing the number of Black subjects in the training dataset.}    
    \label{fig:changing_black_subjects}
\end{figure*}

\begin{figure*}[!hb]
\textbf{\large The effect of changing the level of oversampling in the training dataset}
    \centering
    \begin{subfigure}{0.49\textwidth}
        \centering
        \includegraphics[width=\textwidth]{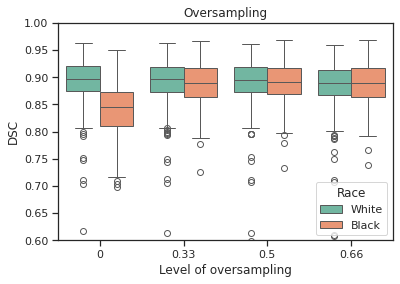}
        \label{fig:oversampling effect}
    \end{subfigure}
    \hfill
    \begin{subfigure}{0.49\textwidth}
        \centering
        \includegraphics[width=\textwidth]{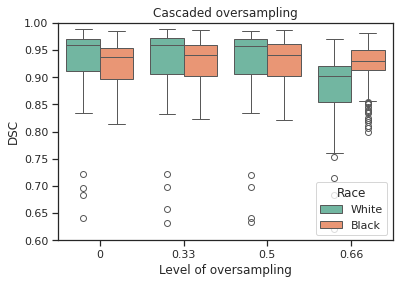}
        \label{fig: oversampling effect cascaded}
    \end{subfigure}
    \caption{The effect of changing level of oversampling of Black subjects in the training dataset.}    
    \label{fig:changing_oversampling}
\end{figure*}

\begin{figure*}[!ht]
    \centering
    \textbf{\large Median DSC against fairness gap for different levels of oversampling}\par\vspace{0.3em}

    \begin{subfigure}{0.49\textwidth}
        \centering
        \includegraphics[width=\textwidth]{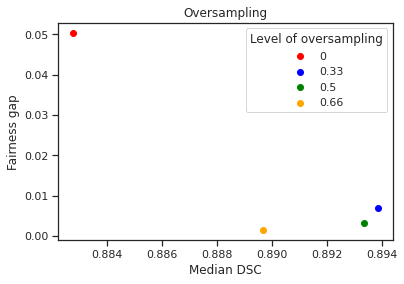}
        \label{fig:performance_vs_fairness oversampling}
    \end{subfigure}
    \hfill
    \begin{subfigure}{0.49\textwidth}
        \centering
        \includegraphics[width=\textwidth]{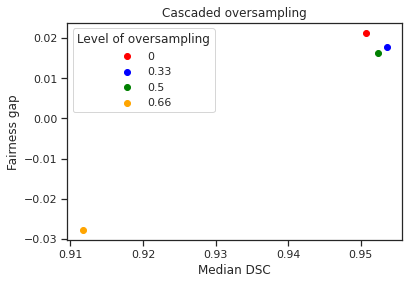}
        \label{fig:performance_vs_fairness cascaded}
    \end{subfigure}
    \caption{Median DSC against fairness gap for different levels of oversampling in experiments using oversampling and cascaded cropping + oversampling. The fairness gap is calculate by subtracting the median DSC for Black subjects from the median DSC for White subjects.}    
    \label{fig:performance_vs_fairness}
\end{figure*}

\end{document}